\documentclass{elsarticle}
\graphicspath{ {./Img/} }
\usepackage{hyperref}
\usepackage{float}
\usepackage{verbatim} 
\usepackage{apalike}

\usepackage{tabularx}
\usepackage[figuresright]{rotating}
\usepackage{graphicx}
\usepackage{lscape}
\usepackage{booktabs}
\usepackage{rotating}

\usepackage{pifont}
\newcommand{\cmark}{\ding{51}}
\newcommand{\xmark}{\ding{55}}

\usepackage{geometry}
\journal{COMPUTER NETWORKS}

\biboptions{authoryear}

\begin{document}
\begin{frontmatter}

\title{Monitoring Fog Computing: a Review, Taxonomy and Open Challenges}

\author[label1]{Breno Costa \corref{cor1}}
\ead{brenogscosta@gmail.com}

\author[label1]{João Bachiega Jr.}
\ead{joao.bachiega.jr@gmail.com}

\author[label1]{Leonardo Rebouças Carvalho }
\ead{leouesb@gmail.com}

\author[label1]{Michel Rosa}
\ead{micheljunioferreira@mail.com}

\author[label1]{Aleteia Araujo}
\ead{aleteia@unb.br}

\address[label1]{Department of Computer Science - University of Brasilia - Brazil}

\cortext[cor1]{Corresponding author.}

\begin{abstract}
Fog computing is a distributed  paradigm that provides computational resources in the users' vicinity. Fog orchestration is a set of functionalities that coordinate the dynamic infrastructure and manage the services to guarantee the Service Level Agreements. Monitoring is an orchestration functionality of prime importance. It is the basis for resource management actions, collecting status of resource and service and delivering updated data to the orchestrator. There are several cloud monitoring solutions and tools, but none of them comply with fog characteristics and challenges. Fog monitoring solutions are scarce, and they may not be prepared to compose an orchestration service. This paper updates the knowledge base about fog monitoring, assessing recent subjects in this context like observability, data standardization and instrumentation domains. We propose a novel taxonomy of fog monitoring solutions, supported by a systematic review of the literature. Fog monitoring proposals are analyzed and categorized by this new taxonomy, offering researchers a comprehensive overview. This work also highlights the main challenges and open research questions
\end{abstract}


\end{frontmatter}

\section{Introduction}
\label{introduction}

Fog computing is a computational paradigm that complements cloud computing, providing computational resources on the network edge, closer to the users. As a distributed infrastructure, fog computing must deal with heterogeneity of network links and processing capacity of its composing nodes \citep{Bonomi2012}.
These characteristics bring complexity to fog management, and it is addressed by the orchestration of services and resources. Orchestration is a management function, composed of several complementary functionalities. It is responsible for dealing with infrastructure dynamicity, for taking timely actions and for assuring that Service Level Agreements (SLAs) are respected \citep{velasquez_service_2017-1}.
There are several proposals of fog service orchestration in the literature, although most of them, only conceptual.  

Monitoring is a functionality of prime importance and it is crucial to properly orchestrate fog services \citep{forti2021lightweight}. It collects updated status information about fog nodes and communication links and send them to the orchestrator. With an updated view of fog infrastructure and service execution, the orchestrator can take proper actions to guarantee the SLAs, e.g., offloading a service to a resource richer node and optimizing service placement according to historic data about node failures \citep{costa2021orchestration}.
Besides the heterogeneity of nodes being monitored, there are other related concerns about frequency, topology, and communication model. There is a trade-off between the frequency of information updates and the overhead to the nodes and to the orchestrator related to generating, transmitting and processing status data. In such a dynamic scenario, adaptability of monitoring parameters can play an important role.
In our previous work \citep{costa2021orchestration}, we did a systematic literature review of fog service orchestration and analyzed 50 proposals. Most of them (40 out of 50) highlighted monitoring as a relevant process, but they frequently assumed that a fog monitoring solution would be available to deliver the information they needed, without presenting either implementation methods or insightful information on the subject. 

Monitoring is not only about reporting availability, i.e. the capacity to answer the question of whether a node or a service is online and working properly. It is also about the capacity to explain why a node or service stopped working properly. The former is achieved by monitoring metrics, e.g. service response time. The latter is achieved by monitoring logs, i.e. unstructured strings of text, and traces, i.e. records of requests made by an user in a service. Metrics, logs and traces form what is called Instrumentation Domains of monitoring \citep{karumuri2021towards}. Different instrumentation domains can be used simultaneously by a monitoring solution to get different perspectives of a service. In such a scenario, there would be more capacity for decision-making on the server-side, but at the cost of increasing the complexity of monitoring, since their specific characteristics (e.g. life-cycle, data volume) would be managed accordingly.
Another emergent concept that is being applied to monitoring microservices is Observability. It is referenced as a superset of monitoring that uses data analytics techniques on the collected monitoring data aiming to shorten the time it takes to know why something is not working as it should \citep{marie2021observability}.



Some works analyzed cloud monitoring solutions and verified that none of them is suitable for use in fog environments  \citep{abderrahim2017holistic, taherizadeh2018monitoring, abreha2021monitoring, battula2019efficient}. As far as we know, only one work \citep{abreha2021monitoring} analyzed fog monitoring solutions and that analysis included only two proposals. In addition, none of these cited works has touched upon observability, instrumentation domains or the monitoring needs of fog service orchestration. 

In order to address these limitations, a comprehensive review of fog monitoring is required. This work analyses fog computing literature and proposes a novel taxonomy of fog monitoring solutions, in line with the ongoing research in this field. By using  this taxonomy, researchers can verify whether a new  monitoring proposal is suitable for use in the context of fog computing. Also, developers of specific fog monitoring solutions can use it as an updated and comprehensive guide to the most important characteristics and features of such a solution.
The main contributions of this work are:
\begin{itemize}
    \item An updated and comprehensive discussion about fog computing monitoring characteristics and composing features;
    \item A fog computing monitoring taxonomy, describing its main domains and categories;
    \item A categorization of fog monitoring solutions found in the literature using the proposed taxonomy;
    \item A discussion about challenges of fog monitoring solutions.
\end{itemize}

The rest of this paper is organized in the following way:  Section \ref{sec:Fog} contextualizes service orchestration in fog computing and related paradigms. Section \ref{Sec:Monitoring} presents requirements and challenges of fog monitoring and describes the review methodology used. Section \ref{Sec:Taxonomy} describes the domains and categories that compose a fog monitoring solution taxonomy. Fog monitoring proposals found in the literature are categorized by the proposed taxonomy in Section \ref{Sec:Evaluation}. Related works are presented and compared with this work in Section \ref{Sec:Related}. The challenges inherent to fog computing monitoring are presented in Section \ref{Sec:Desafios}. Finally, Section \ref{sec:Conclusion} shows conclusions and future work.

\section{Service Orchestration in Fog Computing} 
\label{sec:Fog}
\label{Sec:Orchestration}

This section presents the main characteristics of fog computing and other related distributed paradigms. Also, it contextualizes fog service orchestration showing the relevance of monitoring in this context. 

Fog computing is a distributed computing paradigm that provides resources for computing, storage, and connectivity at the network edge. It can be considered  an extension of cloud computing towards the users' locality. 
It provides computing resources for applications that cannot perform properly with the high latency provided by cloud-only environments. It resides in between the cloud and users, and the cloud will do long-term storage and non-latency-dependent processing \citep{Naha2018}.



The layered (or hierarchical) representation of fog computing is the most widely used approach \citep{Naha2018}. In this context, a three-tiered architecture is the common representation of fog computing environment \citep{Mahmud16}:  IoT Layer, Fog Layer and Cloud Layer. In this architecture, both the Cloud and Fog layers can be implemented through federations. When forwarding a request for resources due to the lack of them in the fog environment, the fog management system can send the request to a Sky Computing, a federation of cloud providers  \citep{2009-Sky-Computing}, which in turn will fulfill the request selecting one available cloud provider. The fog management system itself may also be dealing with multiple federated fog infrastructures. 
The main characteristics of fog computing are \citep{fognist, OpenFog17}: it better deals with low latency needs of services; in contrast to cloud, fog services require widely geo-distributed deployments; on fog, data collection and processing run on different platforms and are delivered by many types of networks; its components should be able to interoperate internally and federate between domains; fog services involve real-time interactions, rather than batch processing, and it supports scalability of needed resources and dynamicity of network and device conditions.

\begin{figure*}[ht!]
	\centering
	\includegraphics[width=0.7\textwidth]{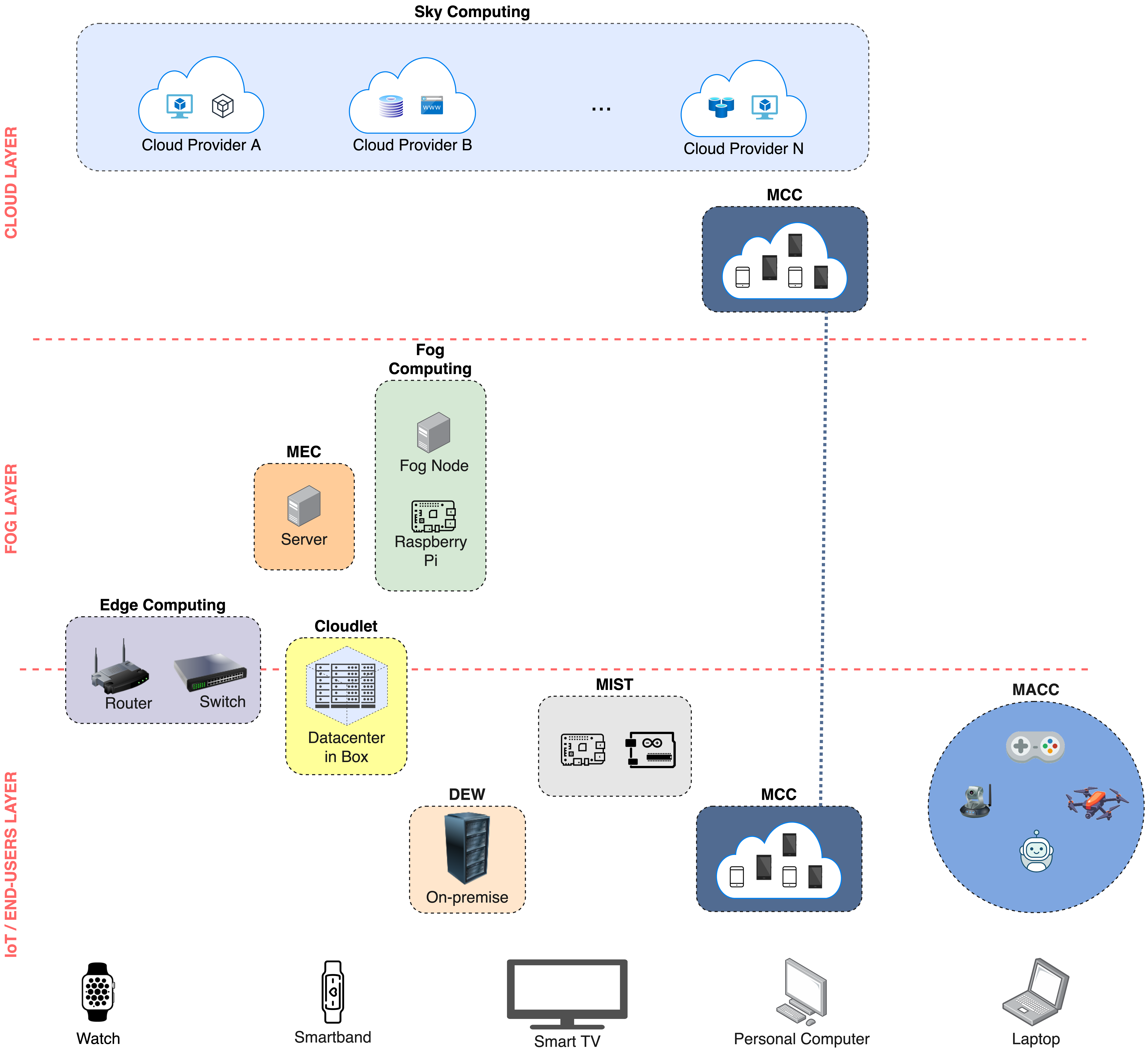}
	\caption{Layered architecture of fog computing and related paradigms.}
	\label{Fig:ParadigmsMerged}
\end{figure*}

\label{subsec:paradigms}
Along with fog computing, there are other distributed paradigms that provide resources on the network edge, but with different architectures and characteristics. Fog computing is often confused with Edge computing, despite key differences between them. Fog has a general-purpose,  multi-layer architecture, while Edge runs specific applications in a fixed logical location. Edge tends to be limited in number of devices \citep{fognist}, whereas fog could scale to a huge number of them.
Besides edge computing, there are other proposed distributed paradigms in literature. Mobile Edge Computing (MEC), Mobile Cloud Computing (MCC), Mobile Ad Hoc Cloud Computing (MACC), Mist Computing, Cloudlet Computing, and Dew Computing are examples of them. In any case, all of them might fall under the same `umbrella’ \citep{stojmenovic2014fog} and eventually solutions proposed for one of them could be applicable to the others under certain conditions and scenarios. In the academy there are publications focused on the presentation and comparison of all these paradigms, such as \citep{Yousefpour2019, Naha2018, Mukherjee2018, bachiega2022paradigms}. Figure \ref{Fig:ParadigmsMerged} shows the layered representation of fog computing and relative position of other related paradigms.

The characteristics of fog computing and related paradigms bring an additional level of uncertainty when compared to cloud computing. This motivates the need for orchestrating fog resources to provide services to users while complying to SLA and Quality of Service (QoS) requirements \citep{kalyvianaki2009resource}.
Orchestration is a management function responsible  for  service life-cycle. To  provide requested  services to the  user and assure  the SLAs, it  must monitor the  underlying infrastructure, react timely to its changes, and comply with privacy and security rules \citep{costa2021orchestration}.

\begin{figure}[h]
	\centering
	\includegraphics[width=0.7\textwidth]{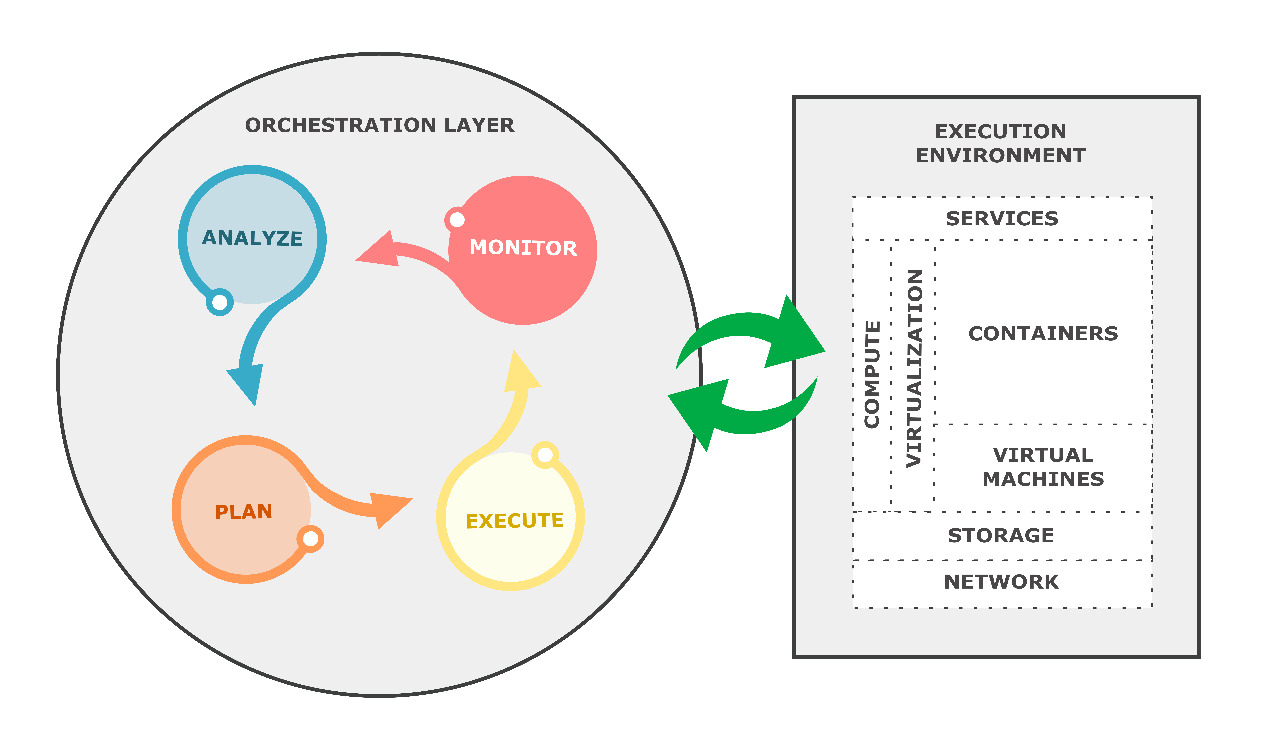}
	\caption{Fog orchestration layer.
	}
	\label{fig:bonomi_architecture}
\end{figure}

Bonomi et al. \citep{bonomi2014fog} proposed a software architecture for running fog services. This architecture is presented in Figure \ref{fig:bonomi_architecture}. It shows a fog orchestration layer, structured as a Monitor-Analise-Plan-Execute (MAPE) control loop that is responsible for providing lyfe-cicle management of fog services in a distributed manner. To illustrate the control loop applied to fog computing, we can start at Monitor phase. In this phase, orchestration must collect updated status about each managed resource and running services. From the Analysis of monitoring data, it can build an updated and comprehensive view of fog environment and Plan the changes needed to maintain the services inside SLAs and QoS limits, as well as providing new requested services. Executing those planned changes will indeed release and allocate proper resources, providing services near end-users.
The authors of \citep{shakarami2021autonomous} used a similar control loop (MAPE-K with K meaning Knowledge) in MEC. They proposed a deep learning approach to improve the decision-making about offloading computation intensive requests aiming guarantee the SLAs. But the Monitoring phase of that loop is responsible to process the requests made by IoT layer and the collection of resource metrics is of responsibility of Knowledge module, where there is a subcomponent called Resource DB. Despite the divergence on the names used, the autonomous computation offloading strategy proposed acts like an orchestrator as depicted in Figure 3, delivering part of the functionalities assigned to Admission Control, Resource Management (offloading), Monitoring (updated Resource DB), Service Management (SLA management) with the data that support them stored in a Repository (Knowledge Module).


There are several challenges that fog orchestration must overcome to guarantee the accomplishment of its goals  \citep{vaquero2019research, velasquez2018fog, jiang2017challenges}: 
\begin{itemize}
    \item \textbf{Churn} - fog resources are inherently volatile. So, orchestration must be aware of resources apparitions and vanishings\citep{jiang2017challenges} as soon as possible to take the proper actions to guarantee the SLAs;
    \item \textbf{Heterogeneity} - not only the devices are heterogeneous. Also, the execution environment can be diverse and occur simultaneously on the same device \citep{Bachiega2021ComputationalPO}, f.i. a fog node with Virtual Machines (VM) and containers concurrently supporting independent services at the same time;   
    \item \textbf{Dynamism} - fog computing is focused on low latency needs. The information about resources and services statuses must be updated to permit an effective decision making about failover and offloading action. Besides, orchestrator must implement its actions within the proper time frame to not impact low latency needs \citep{yi2015fog};
    \item \textbf{Large(r)-scale and Fine(r) Grain} - fog characteristics incur on solutions with smaller code bases and more fragmented state \citep{vaquero2019research}. This increases the complexity of service orchestration, that may have to be aware of a distributed service composition \citep{jiang2017challenges};
    \item \textbf{Security and Privacy} - fog nodes are distributed, resource-restricted and heterogeneous. These characteristics increase the complexity of keeping them secure. The distribution can potentially comprehend several domains and they can have different requirements about the privacy of stored data \citep{jiang2017challenges}. The heterogeneity and resource-restriction may demand different implementation of security enforcement tools and algorithms; 
    \item \textbf{Interoperability} - heterogeneity of resources, access networks and virtualization platforms (VM, container, unikernel \citep{madhavapeddy2014unikernels}) in a fog environment can restrict the interoperability among the orchestrator and the fog nodes; standard communication interfaces, an additional layer for message translation and the implementation of multiple protocols can help overcome this challenge while increasing the requirements a node or the orchestrator must meet to participate;
    \item \textbf{Resilience} - an orchestrator must have a global view about the status of the infrastructure and about the services being provided to the users. By timely instantiating, replicating, and migrating services it can promptly react to the changes and minimize their negative impact in the service availability; 
    \item \textbf{Optimization} - predicting a critical scenario's change and anticipating actions to better deal with it (e.g., a node failure or resource exhaustion), can remarkably impact service's overall performance and availability. But prediction needs Machine Learning (ML) algorithms and datasets that describe the previous behavior of subjects of concern. The amount of data being stored and transmitted in the network to the orchestrator and the proper ML algorithms to be used are challenges to overcome towards the use of optimization;   
    \item \textbf{Authentication, Access, and Account (AAA)} - the interaction between the orchestrator and each fog node can be modeled into the steps of an AAA framework. Authentication queries node's credentials and validate whether it is allowed to participate on that fog infrastructure or not. After node's authentication, the orchestrator can access local execution environment, run commands and services, and collect node's metrics values. Finally, the account records resource usage and this data can be used to reward node's owner and/or bill the users who consumed the services;
    \item \textbf{Fine-grained Locality} - orchestration must manage fog nodes' resources in a way that does not overburden them. 
\end{itemize}

 To overcome the aforementioned challenges, fog service orchestration delivers several complementary functionalities
 \citep{velasquez_service_2017-1, wen_fog_2017-1}: 
 
 \begin{itemize}
     
     \item Admission Control of Incoming Requests - the interface with the end-user. Receives the requests, assesses the requester's credentials and decides where it will be served (fog/cloud).
     
     \item Service Management - manages service life-cycle, i.e., service registration, service images to the different virtualization platforms, service constraints, and requirements. Uses monitoring data to verify the need for taking actions aiming to guarantee the SLAs.

     \item Resource Management - manages resource life-cycle, i.e., discovers new nodes, allocates resources to fulfill accepted requests or a need for offloading, deallocate resources; Monitoring data is responsible for maintain updated the resource inventory. 

     \item Monitoring - updates status of availability and usage of resources and services and manage also logs and traces.
     
     \item Optimization - processes available data with the use of algorithms and techniques to minimize some metrics and/or maximize others; e.g., the authors of \citep{shahidinejad2021resource} proposed an optimization process using an evolutionary algorithm and ML techniques to clusterize services according to their SLAs, improving resource provisioning. Its benefits are a potential decrease in cost and response time and an increase in CPU use an in elasticity.
     
     \item Communication Management - uses protocols and standards to cope with heterogeneity of nodes and communication links of a fog infrastructure.
     
     \item Node Agent - a local agent that manages the execution environment of a fog node, realize the actions required by service manager, e.g., replicate a service, download a new service image, and collects monitoring data.
     
     \item Security - is responsible for the enforcement of security and privacy policies.
 \end{itemize}

\begin{figure*}[ht]
	\centering
	\includegraphics[width=\textwidth]{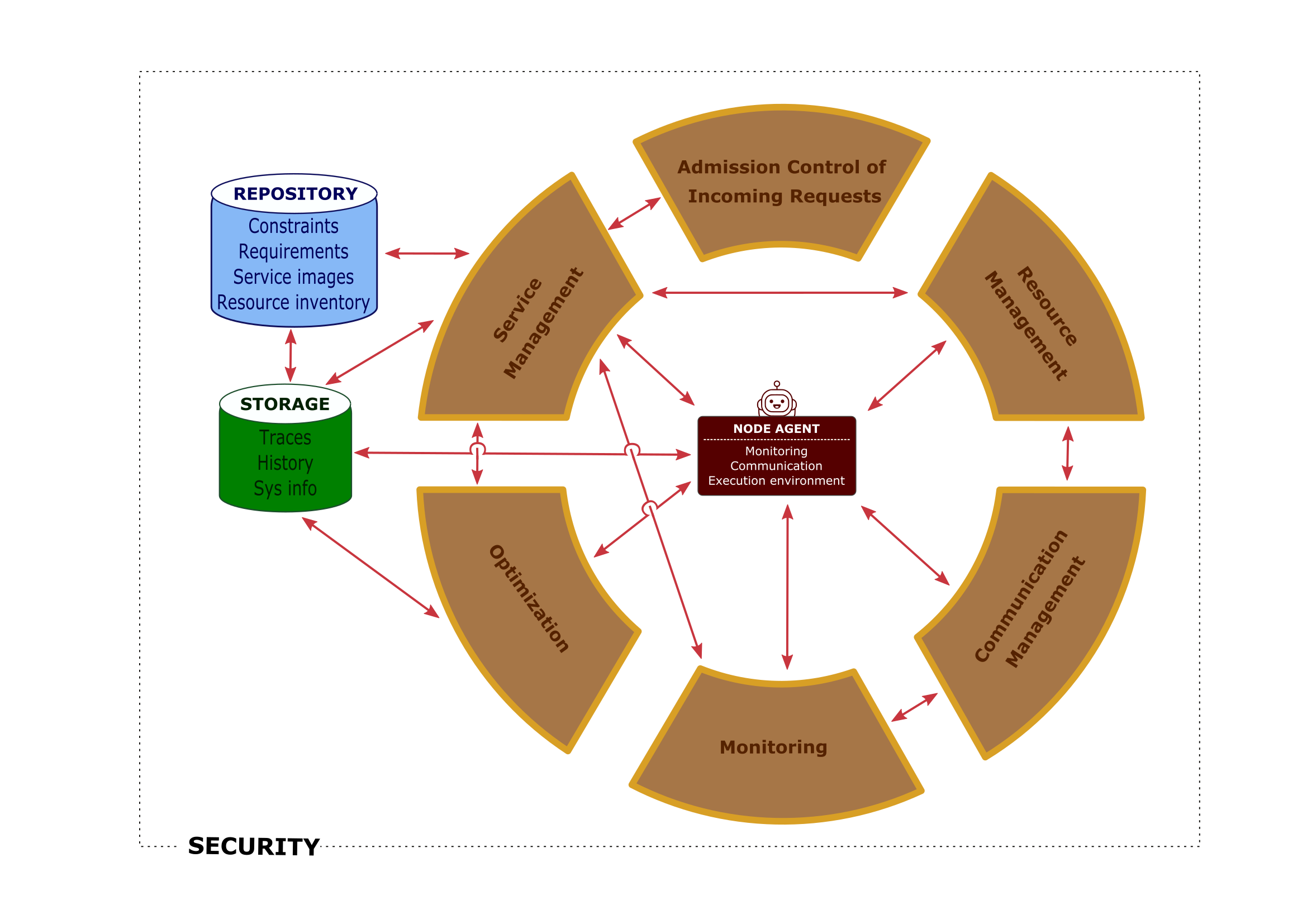}
	\caption{A generic fog computing orchestration architecture \citep{costa2021orchestration}.}
	\label{fig:architecture}
\end{figure*}

 Figure \ref{fig:architecture} depicts  the interactions those functionalities have by means of a generic architecture created from the analysis and consolidation of 50 fog orchestration works available in the literature \cite{costa2021orchestration}. 
 

\section{Monitoring Fog Computing} 
\label{Sec:Monitoring}

This section describes the main characteristics of monitoring (e.g. sub-processes, system components) and introduces emergent concepts in this field (e.g. observability, instrumentation domains), gathering state-of-the-art knowledge in the area of fog computing. It also provides an analysis of the role of monitoring in the context of fog service orchestration, regarding  requirements and challenges of fog monitoring. Finally, the section presents the methodology used in the systematic literature review.

Considering the characteristics of fog, systems running in this environment will have a high distribution of their components, variable and unpredictable load, caused by the heterogeneity of devices, communication links, and failures. This scenario makes it challenging to predict how these systems will behave over time \citep{arpaci2018cloud}. Infrastructure monitoring is the basis to support several goals: make efficient use of resources, measure resource and service performance, generate accurate bills \citep{syed2017cloud} and implement fault tolerance processes. The previous section showed the importance of monitoring features for supporting proper orchestration of resources and services in a fog computing environment. 

A monitoring service can be structured as a composition of three different functions: 1. observation of monitored resources and services; 2. data processing; and 3. data exposition \citep{abderrahim2017holistic, brandon2018fmone}. Observation means the acquisition of updated statuses of resource usage (e.g., CPU load and latency) or service performance (e.g., response time). Processing is related to the necessary adjustments and transformation required on data, such as filtering and aggregation, creation and management of events, and notifications derived from pre-configured rules and thresholds. Exposition is related to where the generated data is stored (e.g., in a local database, JSON files) and how it can be accessed by a management system (e.g., visualization through dashboards, other functionalities consuming  the data directly).

\subsection{Instrumentation Domains}
\label{subsec:instrumentation}
Monitoring has three instrumentation domains: metrics, logs, and traces \citep{karumuri2021towards}. Each domain has its own characteristics and supports different decision-making processes, which can occur before, during and after the start of data collection of a particular monitored object. A metric is a measurement at a particular point in time. It is represented by a name, a value (the measure), a timestamp, and other associated (optional) context data. A log is a collection of unstructured or semi-structured strings. They bring detailed information and additional context. A trace is the representation of a single operation, e.g., a request from an user inside a service, showing the entire execution path taken from the beginning to the end of the request \citep{karumuri2021towards}.

More recently OpenMetrics \citep{openMetrics} has been published, and it is intended to be primarily a standardized format for metrics exposure, independent of any specific transport. Currently, there are dozens of agents that can export metrics using this standard and several of them have characteristics suitable for use in fog \citep{openMetrics}. OpenTelemetry \citep{openTelemetry} is the standard for representing and communicating traces. The use of such data format standards facilitates the composition of solutions based on several independent monitoring elements. On the other hand, OpenTelemetry suffers from a lack of tools for managing monitoring data and still does not allow automated analysis due to not having a strict enough specification \citep{bento2021automated}.

Using terminology created in the software testing domain, monitoring can be divided into black-box and white-box. Black-box is the monitoring done from the public interface of the object being monitored and aims to answer whether or not the object is available/working, i.e., it identifies if there is a problem. White-box is based on the collection of detailed information on the functioning of the object's internal processes and aims to answer why the object is not available or functioning properly, i.e., it allows a root cause analysis \citep{ewaschuk2016monitoring}. Metrics are more related to black-box monitoring, while both logs and traces are more related to white-box monitoring.

There are dozens of metrics that can be collected from the monitored subject, no matter if it is a physical or virtual device (e.g., \% of CPU usage, \% of free memory, etc.),  a container management system like Kubernetes \citep{noauthor_kubernetes_nodate} (number of containers, number of requests by container \citep{ifrah2021getting}) or a (micro)service running on a fog node (e.g., response time etc.). The literature defines the golden signals of monitoring. It is a minimal set of telemetry data (i.e., instrumentation data) that provides essential information for proper black-box monitoring of infrastructure, platforms, and services. They are: latency, traffic, errors and saturation \citep{ewaschuk2016monitoring}. 

The choice of which metrics should be collected to better represent an asset's status is a challenging issue in monitoring context in fog computing. Increasing the number of metrics can mean more overhead in the collection and more data to transmit, analyze and store. On the other hand, valuable information must flow in the monitoring process, allowing proper decision making to happen and helping other orchestration functionalities to reach their goals. There should be flexibility in selecting the set of metrics that are sufficient to accurately inform monitored object status and to make decisions timely, according to fog computing requirements.

\subsection{Observability}

Observability is a term borrowed from control theory. In computer science, it is defined as a characteristic of software and systems related to the information they generate that allows them to be monitored and understood more comprehensively, including at runtime. In addition to a simple black-box monitoring, Observability can provide a greater understanding of the correctness and performance of services. One of its goals is to shorten the time it takes to know why something is not working as it should. It is an inherently data-intensive and time-sensitive process \citep{karumuri2021towards}.

Observability is an emergent concept that has been used to reference advanced monitoring functions in the context of microservice-based applications. Observability is sometimes considered to be a superset of monitoring, since it aims to fulfill the same purposes and additional ones, by extending monitoring concept and applying data analytics techniques on the monitoring data \citep{marie2019demonstration}. Observability is more related to white-box monitoring, and to the generation and consumption of traces as the main basis of information and decision-making. 

In this work, we will use only the term ``monitoring'' to name the process of collecting information (metrics, logs and traces) from monitored objects (hardware, virtualized environments and services), take actions based on the collected information, and store it for further analysis or for long retention.


\subsection{Monitoring System's Components}
The monitored subject needs to provide access to its metric values. This is usually implemented using operating system calls made by a local monitoring agent that collects the data in a predefined recurrent period and make them available for processing. The same agent, or another specialized process, can regularly check if the data collected reach some predefined threshold and, if it does, take proper monitoring actions, like creating an event or notifying the monitoring personnel. Finally, the monitoring server is accountable for storing monitoring data and make them available for other management processes. Although described as three independent components, they can also be located in the same execution environment, implemented as different functions of the same service.

\begin{figure*}[ht]
	\centering
	\includegraphics[width=1.1\textwidth]{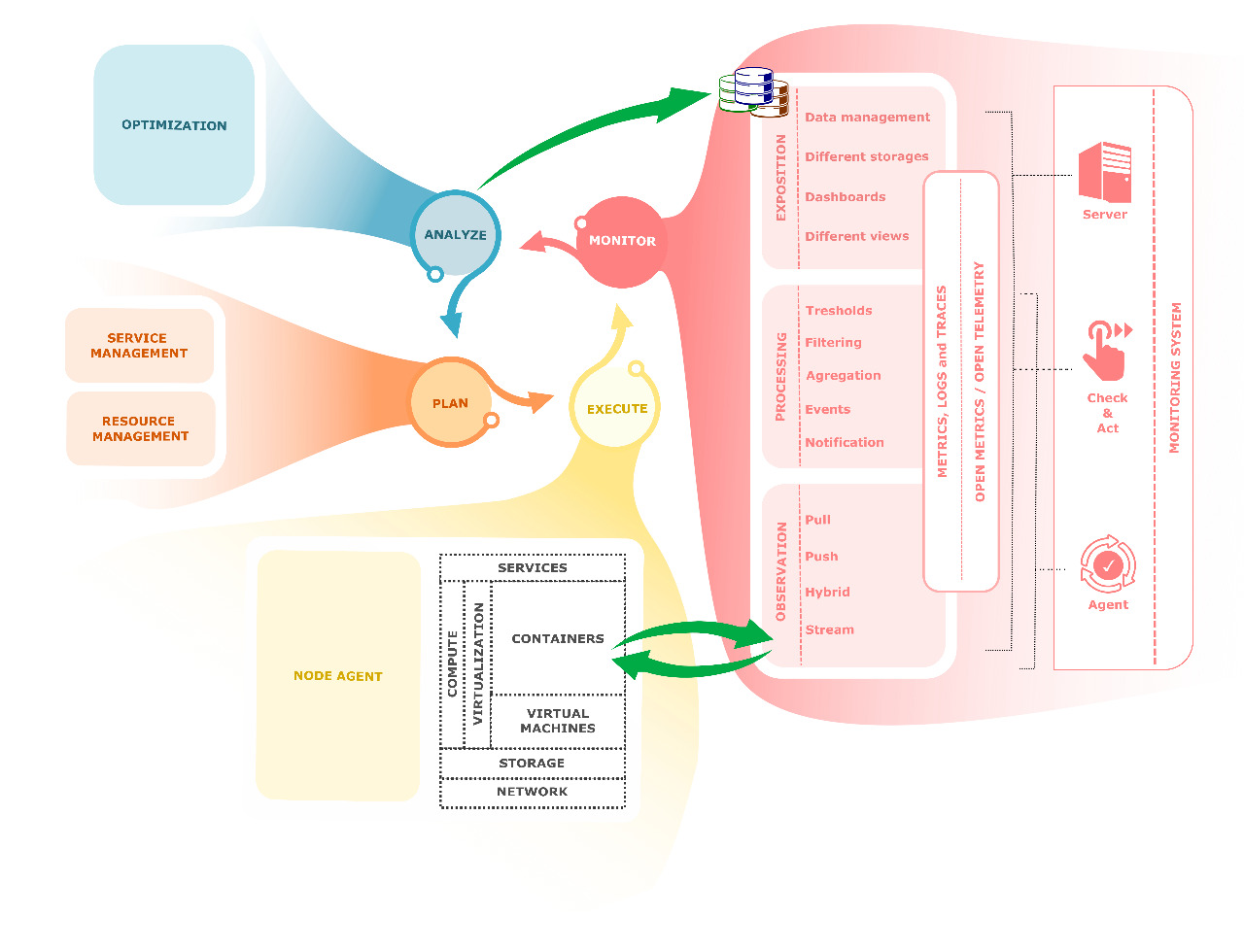}
	\caption{Fog orchestration layer detailing monitoring functions, instrumentation domains, and system's components.}
	\label{fig:integratedView}
\end{figure*}

Figure \ref{fig:integratedView} shows an in-depth view of the Fog Orchestration layer already shown in Figure \ref{fig:bonomi_architecture}. In Figure  \ref{fig:integratedView}, the Monitor phase is detailed to expose monitoring functions, monitoring system components, and the instrumentation domains. The orchestration's functionalities (Section \ref{Sec:Orchestration}) that are responsible for implementing the phases Analyze, Plan, and Execute, are also shown.

\subsection{Fog Monitoring Requirements and Challenges}
\label{subsec:requirements}
To properly deal with fog characteristics, e.g., resource-restricted, and heterogeneous devices, variety and instability of connections, fog monitoring solutions must meet novel requirements that are not available on cloud monitoring solutions in place \citep{abderrahim2017holistic}, thus requesting specific approaches. According to Abderrahim et al. \citep{abderrahim2017holistic}, the solutions for monitoring fog infrastructures must have the following properties:
\begin{itemize}
    \item Scalability - to deal with the increased  number of fog nodes;
    \item Resilience to Node Apparitions/Removals - mobility is a potential characteristic of a fog node;
	\item Resilience to Network Changes/Failures - since this is a fog characteristic;
	\item Modularity - there must be room for adaptation or parameterization as there are different scenarios of service execution in fog environments according to each specific use case, e.g., resource capacity and network stability can be so much different among Industrial IoT (IIoT) and vehicular traffic management use cases;
	\item Locality - the monitoring must be as nearest as possible to the resources and services being monitored.
\end{itemize}

According to Taherizadeh et al. \citep{taherizadeh2018monitoring}, monitoring solutions within fog computing have several challenges:
\begin{enumerate}
\item Data Management - collection, processing, and transmission of instrumentation data can overburden the network in a large scale environment. There is little research on the storage and management of logs and traces \citep{karumuri2021towards}. 
\item Coordinated Decentralization - when the control topology is hierarchical or distributed there is a risk of desynchronization of management actions and loss of effort and time to resync the distributed components (managing consensus and synchronizing replicated data) \citep{taherizadeh2018monitoring}; Centralized topologies are already addressed by enough research attention from the academia and there are many solutions proposed;
\item Fault Tolerance - the service could continue to operate under a faulty event \citep{chang2014scalable}, e.g., when the node lost the connection to the orchestrator, but the requester is consuming the service normally. Off-line detection and recovery mechanisms should be necessary to reintegrate the node to the infrastructure and acquire the generated data;
\item Mobility Management - a moving end-user device can have a variation on the network parameters of the connection to the fog node, rapidly changing service's Quality of Experience (QoE) \citep{ahmed2016survey}; Depending on the configurations of data collection, this information can reach decision-making module with a delay that can impede timely actions to preserve the SLA and QoE inside the range;
\item Scalability and Resource Availability on the Edge - although there is a resource restriction on fog nodes, when a service is executed, it should accommodate certain demand increase at the risk of causing service unavailability due to the lack of resources \citep{ahmed2016survey};
\item Prior Knowledge - it is necessary previous knowledge about underlying infrastructure and distribution of service components to guarantee a node comply to QoS requirements \citep{xiao2017self};
\item Interoperability and Avoiding Vendor Lock-in: vendor lock-in is a cloud disadvantage \citep{toosi2014interconnected}. EdgeX Foundry \citep{edgexfoundry}, Open Edge Computing \citep{openedgecomputing} and OpenFog RA \citep{OpenFog17} are projects that aim to standardize a framework for fog  computing;
\item Optimal Resource Scheduling Among Fog Nodes - Resource scheduling should be aware of dynamicity of execution environment concerning user's mobility, and variation on connection QoS. It should guarantee fast responses under these conditions \citep{lee2015internet};
\item Proactive Computing - to anticipate critical events, and trigger actions and decisions to deal with them proactively, time constraints must be considered, and this demands large amounts of historical data \citep{fournier2015extending};
\item Replication of Services - service replication is a strategy to increase service performance, availability, and fault tolerance. But it also increases the management complexity, demanding synchronization checks \citep{farris2017optimizing};
\item Container Security - the use of containers as execution environments, although they are considered adequate to fog computing, can pose new security threats \citep{sultan2019container}; along with this risk, there are plenty of suitable tools and techniques that can lower the attack surface of containers.
\end{enumerate}

Together, these monitoring requirements and challenges present another perspective for highlighting the main differences between cloud monitoring and fog monitoring. Heterogeneity of fog nodes and larger distribution require that the fog monitoring solution be platform agnostic, interoperate with a variety of virtualization environments and be prepared to bursts on the number of devices and in the data volume. A way to achieve this is being modular and scalable. The expected low-latency of fog applications require that collected information be available timely to decision-making, but a balance is needed to tackle resource-restriction of fog nodes, unstable connections and device mobility. It is fundamental to be closer to the items being monitored, provide different strategies to collect the data (bulk collection, push vs pull, adaptive rate, adaptive metric set etc), and be resilient by means of fault tolerance management.

\subsection{Monitoring as a Function that Makes up Fog Orchestration}
\label{Subsec:MonitOrch}
Monitoring is a functionality of prime importance in fog orchestration, responsible for collecting instrumentation data, e.g., CPU usage, service response time etc., and to maintain updated the status of resources and services. This information can be useful for decision making about service placement and offloading, scalability issues, SLA and QoS management, actions and strategies related to other complementary fog computing orchestration functionalities seen on Figure \ref{fig:architecture}. Monitoring can be seen as a glue that bounds orchestration's functionalities together.

Despite being an important functionality, an orchestrator's monitoring module must focus on its specific goals: collect data on resource usage and performance, generate events when pre-configured thresholds are reached, transmit the events, notifications and collected data to a decision-maker module. Being focused means monitoring solutions that address the needs of a fog orchestrator should not implement functions associated with other orchestration's functionalities. There would be no value if the collected monitoring  data and generated events were not used for decision making about resource and service life-cycle management. But Service Management and Resource Management modules are the ones accountable for making these decisions. And they do so using the orchestrator's broader view of fog infrastructure and services delivered, besides end-user relationship with fog and cloud provider. 

Orchestration demands that complementary functionalities act together, but each one doing its specific job. Several fog orchestration challenges (seen on Section \ref{Sec:Orchestration}) like heterogeneity, dynamism, resilience, AAA, and fine-grained locality point out to lightweight implementations of each module aiming that orchestration overhead be as negligible as possible in fog environment. Monitoring solutions (and this is valid also to other functionalities of an orchestrator) that incorporate tools and functions of other modules, although could be used as a stand-alone fog service, are not proper to compose an orchestrator due to potential resource-wasting and high overhead of having replicated functionalities.

In previous subsection, only the first three monitoring challenges (data management, coordinated decentralization and fault tolerance) presented by Taherizadeh et al. \citep{taherizadeh2018monitoring} are specific of a monitoring module that composes a fog orchestrator. The remaining challenges are primary concern of other functionalities as Service Management (Items 4, 5, 6 and 10), Resource Management (Items 5, 6 and 8), Optimization (Items 4 and 9), Security (Item 11), Communication Management (Item 7).

For example, in the challenge of Mobility Management (Item 4), monitoring is collecting and recording infrastructure and service metrics in the same way that if the end-user was still. But, if one of the metrics collected is directly related to QoE, e.g., service response time, as soon as this information is available to the orchestrator, Service Management can verify if response time is within the agreed range and act if it is not. The action could be to replicate the service closer to the end-user's current location. To accomplish this, it can call Resource Manager and pass, as parameters, service information and requirements. Resource Manager will allocate a new fog node (or return informing none is available and forward the requisition to the cloud) looking into resource inventory and place the service replica. So, Mobility Management is not a specific fog monitoring challenge, but a fog orchestration one. 

This specialization is important not only to save resources and to comply with low latency needs, but also because fog can add value to several use cases with different requirements of connectivity, mobility etc. To be useful in different scenarios, a fog orchestrator should be implemented on a modular way and this property \citep{abderrahim2017holistic} should be expanded to its functionalities. Each orchestration functionality should permit parameterization and adaptation, so it can meet the requirements of specific use case in place. The more focused a monitoring solution is, taking care of only the monitoring process, events, and data, the easier it is to switch it, when needed, to another that could be more appropriate in a different scenario (e.g., inside a proprietary device).

\subsection{Research Selection Method}
\label{Method}
This subsection describes the method used to systematically review the literature on Fog Monitoring. The review was inspired by the works of  \citep{kitchenham2009systematic} and \citep{petersen2008systematic}.The steps taken were: 1. define the research questions (RQ); 2. choose the research databases; 3.create a search string made of relevant keywords; 4.gather all results; 5.apply inclusion and exclusion criteria; 6.filter the studies based on keywords, title and abstract, and 7. read and analyze remaining studies.

We defined the following research questions to guide the systematic review: RQ1 - What are the relevant characteristics of a suitable fog monitoring solution?. The answer to this question will be consolidated as a novel taxonomy and it will help the researchers to identify what relevant features a fog monitoring solution should have; RQ2 - What monitoring solutions are prepared to compose a fog service orchestrator? The answer to this question will come using the taxonomy to categorize the state-of-the-art proposals selected by this systematic review of the literature; RQ3 - What are the challenges that still need attention from the academia? The answer to this question can be used by researchers as a guide for future works in this area.

For this article, Scopus\footnote{scopus.com}, Web of Science\footnote{webofknowledge.com}, ACM Digital Library\footnote{dl.acm.org} and IEEE Xplore Library\footnote{ieeexplore.ieee.org} databases were used as research sources. The basic search string created was ``(Fog OR Edge) AND (Monitor* OR Observability)''. Complementary searches were made using other distributed paradigms' names (e.g. MEC, Cloudlet etc, as listed in Section \ref{sec:Fog}). 
Inclusion criteria were: peer-reviewed primary works; written in English; publication date starting in 2012 (year of first fog computing publication \citep{Bonomi2012}; and works that present solutions, architectural models, techniques or methods applied to monitoring in fog computing. After running the searches and gathering all the results, the duplicates were removed and the 791 remaining studies were filtered based on keywords, titles and abstracts. The resultant set was composed of 75 works. Finally, after reading and analyzing the full text of the remaining studies, we selected the 10 works that are detailed in this paper and by this means collected the necessary knowledge to answer to the research questions.

\section{Taxonomy of Monitoring Characteristics in a Fog Orchestration Scenario} 
\label{Sec:Taxonomy}

This section presents a novel taxonomy of fog monitoring solutions, created from the systematic review of the literature. The taxonomy consolidates domains and categories that are relevant in a state-of-the-art fog monitoring solution. The content provided by this section answer the first research question:  ``RQ1 - What are the relevant characteristics of a suitable fog monitoring solution?''. 

Several works proposed taxonomies of cloud computing monitoring solutions \citep{aceto2013cloud, ward2014observing, syed2017cloud, da2019survey}. They classified available cloud monitoring tools using taxonomies.
Using the process described by Usman et al. \citep{usman2017taxonomies}, we created a taxonomy to categorize fog computing monitoring solutions. Despite the differences between cloud and fog, some domains and categories of cloud monitoring solutions are applicable in fog monitoring solutions, e.g., topology and frequency of data transmission, while others are not applicable (e.g., types of cloud: private/public). But even those domains that are applicable need to be reviewed, and eventually adapted, to reflect the novel requirements and scenarios of fog computing.

\begin{figure*}[ht]
	\centering
	\includegraphics[width=1\textwidth]{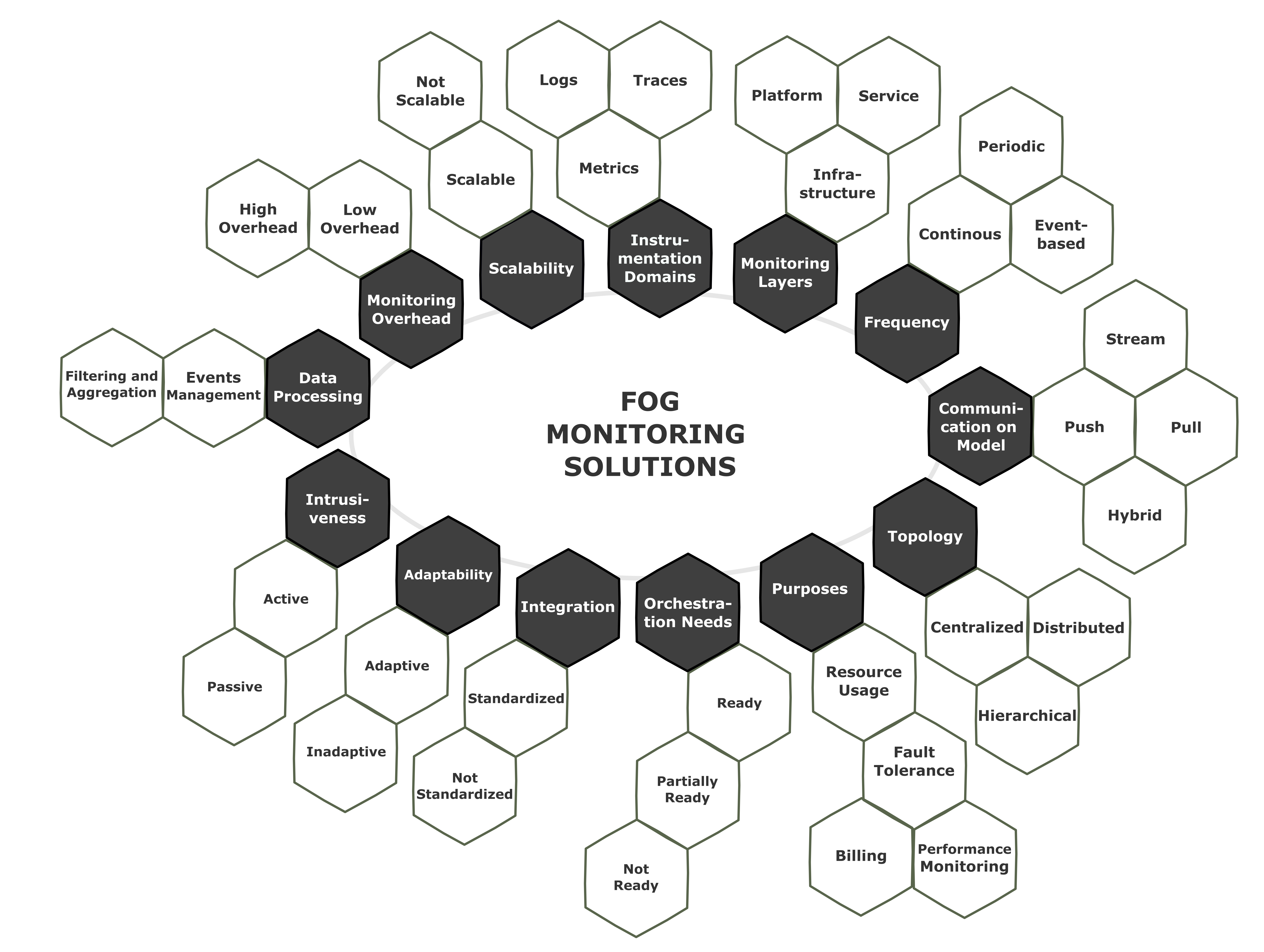}
	\caption{Taxonomy of a fog monitoring solution.}
	\label{fig:taxonomy}
\end{figure*}

We analyzed cloud taxonomies and reviewed the literature about monitoring fog environments and services. Based on the challenges of fog computing orchestration (Section \ref{Sec:Orchestration}), on the characteristics of fog monitoring solutions (Section \ref{Sec:Monitoring}), and on several works selected from the literature, we identified domains and categories that are relevant in this context. They are summarized in Figure \ref{fig:taxonomy} and a detailed description of them is presented in the following subsections.

\subsection{Monitoring Purposes}
This domain defines the goals of a fog monitoring solution. The most frequent ones found in the literature are track resource/service usage and performance monitoring. From the data collected to reach these goals, fault tolerance and billing processes can also be supported \citep{battula2019efficient}.

A fog orchestrator, or other management system, must have updated information about its managed resources to make informed decisions. Each resource type (e.g., CPU and network) will have a specific set of metrics that inform their status: e.g., percentage of free CPU and percentage  of package loss, respectively. According to a static or dynamic configuration, a monitoring solution must collect metric values that are relevant to the other management processes, store them locally, when there is enough room, and transmit them timely to persistent storage and analysis.  

Fog computing is a more distributed paradigm, and its nodes and communication channels (physical and virtual) are potentially resource-restricted and unstable, due to heterogeneity and mobility. In such a risky scenario, performance monitoring can leverage SLA management made by the fog orchestrator. Performance monitoring can be done in different levels or service models and each of them can have its own performance indicators. Performance of a communication channel can be measured by the throughput in bytes per second. Performance of a database can be measured by the throughput in transactions per second, and performance of a service can be measured by response time. 

Another role of a fog monitoring solution is the support for fault tolerance processes. To verify a node availability, a heartbeat message can be sent from the node to the orchestrator in a timely fashion. After a given period without such message, the node is considered offline, and the service orchestrator can make decisions about allocating similar available resources, about migrating services to other nodes, and communicating the users. A fog monitoring solution can be accountable by implementing this keep-alive process and by generating the events of unavailability to the management system.

Although the business model of fog computing is not defined yet, the records of resource usage and allocation are the basis to describe a user's consumption and generate accurate and verifiable bills when needed \citep{syed2017cloud}. 
Billing records can be reported in a different granularity when compared to resource usage and performance. 

A fog monitoring solution that has several purposes could need different communication models (see Subsection \ref{subsec:CommModels}) and parameterization to reach them properly.

\subsection{Monitoring Topology}
Monitoring topology describes how the monitoring system is structured in terms of distribution of its components and data flow. Masip et al. \citep{masip2020collaborative} described three control topologies that can be used in fog environments and were adapted to describe monitoring topologies: centralized, with only one monitoring server; hierarchical, where a set of resources/nodes have a local monitoring server and these servers act somehow on the data (filtering, storing, etc.) and collaborate among them in a pre-defined way; distributed, where a monitoring component is localized on each fog node and all components interact to share their view of monitored resources and together maintain the view of the whole environment updated. We have adopted `hierarchical` instead of `decentralized', as named by Masip et al. \citep{masip2020collaborative},  because it seems more meaningful in this context.

In a centralized monitoring topology, there is only one monitoring server. This server receives all the instrumentation data sent by the agents deployed on monitored nodes, or the server itself queries each node about the data of interest. 
This topology is easier to implement, making the monitoring agents simpler, but has some disadvantages. Firstly, there is the Single Point of Failure (SPOF) issue, where a server failure may interrupt monitoring updates of the whole fog environment and harm decision making; secondly, the server must run in a resource-rich node to cope with data flow and storing. A possible solution to this issue is to put the server in a cluster of fog nodes or in the cloud and use different communication models and data priority policies to cope with fog-cloud communication latency; lastly, server network channels can become overloaded with monitoring data flow.

In a hierarchical monitoring topology, there is at least one more additional data staging layer between the monitored node and the monitoring server. The nodes are categorized by locality, and assigned to a local monitoring server. This intermediary component may function to those nodes as a centralized server, to where all monitoring data is transferred and stored by a longer time than the monitored node could.  Also, it is possible to filter and aggregate the data before sending the data to support decision making. But this local server may not be the final destination of monitored data. The monitoring server can be an independent component in the system, receiving and managing a high-level view of monitoring data and integrating to the service orchestrator. Otherwise, monitoring server can be implemented as a Peer-to-Peer (P2P) network of local servers and the hierarchical topology will function as a hybrid topology between centralized and distributed ones.

In a distributed monitoring topology, the monitoring server is implemented as a P2P network of components that are distributed in all nodes. The collected monitoring data must be shared (replicated) with all the peers that need them and eventually with the service orchestrator for decision making about resource and service management. 
This topology is referred by Abderrahim \citep{abderrahim2017holistic} as being the best option for fog environments. It has the benefits of overcoming issues of monitoring server found on Centralized topology (f.i. SPOF, runs only on resource rich nodes, network congestion). Nevertheless, distribute this relevant function by some unreliable nodes can lead to outdated data on resource statuses. Also, the bigger the P2P network of distributed servers, the higher the risk of desynchronization of replicated monitoring data.


\subsection{Communication Model}
\label{subsec:CommModels}
A fog monitoring solution should collect instrumentation data from a monitored node, resource or service and make this data available timely to the service orchestrator to support decision making and data analysis. The data can be sent by a monitoring agent to the monitoring server periodically, and this category of communication model is called Push. The monitoring server can ask for data in an event-driven manner, and this is called Pull. The mix of both models create the Hybrid category. Lastly, the Stream model describes a continuous data flow between the agent and the server. 


In Push model, the monitoring agent is responsible for initiate the transmission of collected data to the monitoring server. 
The agent needs to know previously the server address. This information is sent to the agent in the bootstrap process and can be updated while it is running, based in server's events (e.g., server is  overloaded) or based in monitored node's events (e.g., running out of energy).



In the Pull model, the monitoring server is responsible for initiate data transmission and needs to request the data to the monitored nodes. Only after a proper request, the data is sent by the agent.
In this model, the server can select the information needed at the moment, sending parameters inside the request. This could help to implement a strategy where the server asks for high priority monitoring info when it is overloaded and asks for bulk info when it can cope with the burden. 



A Hybrid model is a more flexible way of dealing with dynamic situations and with heterogeneous monitored infrastructure and services. Based on resource capacity of the nodes, e.g., nodes running out of data storage can use push model as a way of not losing data. Nodes that are resource-richer can use push model only when selected metrics reach a predefined threshold, e.g., every 10\% increase or decrease of free CPU. In other situations these high capacity nodes can store monitored data and wait for a bulk request sent by monitoring server. There are other combinations of Pull and Push models that can be configured to satisfy specific requirements or scenarios where the use of only one model is not sufficient.


In a Stream model, the agent creates a data flow with the server and transmit monitored data continuously. This is a model appropriated to transmitting small volume of priority data, as heartbeat messages and high priority notifications.

\subsection{Monitoring Frequency}

On orchestration, an effective decision-making is supported by updated information about resources and services. The higher the frequency of updates, e.g., every change in a metric value being delivered immediately to the monitoring server, the lower the risk of dealing with outdated information. But a high frequency of updates can cause computational overhead and network congestion mainly when dealing with a high number of monitored devices. The monitoring frequency must be balanced with computational capacity and network overhead.
There are different frequency approaches for monitoring:
\begin{itemize}
    \item Continuous - once started, monitoring data flows to the server continuously;
    \item Periodic - where a recurrent period can be configured and data is sent at every time instance; 
    \item Event Based - where data exchange is triggered by the detection of an event on the node or by answering an explicit request by the server.
\end{itemize}


\subsection{Monitored Layers}

A fog environment is composed mainly of fog nodes, physical or virtual devices that execute fog services \citep{Bachiega2021ComputationalPO}. To provide requested services to the end-users, a fog orchestrator should manage the resources and service distribution. 
Due to the heterogeneity of fog nodes, a service can run on their bare metal, or on the virtualization platform available on the node (VM, container, unikernel), considering that Service Management (Section \ref{sec:Fog} has specific service image to each of them.

A monitoring system should collect metric values from several layers inside the nodes, each one potentially demanding specific probes and generating specific events and notifications as they are independent from each other.
Metrics associated to the hardware, such as free CPU and RAM, are normally collected from the operating system, although they also can be provided by the virtualization software (Docker, f.i.). Metrics associated to the virtualization platform may also be of interest, as well as, metrics associated with the service, such as response time.
Monitoring agent on the node (or outside the node) must collect these metric values by using specific probes and using configured protocols and ports. This domain indicates to which layers a monitoring solution is prepared to collect metric values: infrastructure, platform or service layers.

\subsection{Instrumentation Domains}
As seen in Section \ref{subsec:instrumentation}, there are three instrumentation domains for monitoring distributed systems: metrics, logs and traces. The use of metrics is well-known and there is support for it in several monitoring solutions. It allows a black-box monitoring, collecting data from the operating system, container management solution or other virtualization environment, without the need to modify the system being monitored. Metrics are lightweight when compared to logs and traces. Due to these characteristics, metrics are well fitted for fog environments, made of resource-restricted and potentially mobile devices, and connected by unstable communication channels.

On the other hand, logs and traces permit analysis from system's internals, enabling discovery of the causes of bad service performance, and predicting future issues that could be caused by a current misbehavior.
Logs and traces can make use of the data flow management established to cope with metrics data. However, due to the need for a higher resource availability, its impact on the service's SLA must be evaluated prior its use and a dynamic way of turning it on and off should be provided.

\subsection{Data Processing}

A fog monitoring solution should process collected data (e.g., filtering, aggregating, transforming) as part of processing function \citep{abderrahim2017holistic}.
There is a trade-off in this domain. As much data processing features are available in a monitoring solution, a lower volume of monitoring data may flow to the monitoring server, reducing network load. Nevertheless, to run these features properly a resource-richer node must be available.
In the category of events management, collected data could trigger an event and the event can generate actions on the node itself (e.g., stopping collection when exhausting the CPU) or on the monitoring server (e.g., sending a notification). 

According to data management strategy in place, as soon as the data is collected from the data source (e.g., an end device, a platform like kubernetes or a deployed service) it can be filtered, aggregated or suffer other kind of modification before it is stored and transmitted to the monitoring server.

Some fine-grained metric values can be stored in-place for a short period of time. This can allow the generation of alerts for specific pre-configured situations and the recovering of detailed data when needed in that time window. Also, device's resource limits can be respected.  A coarse-grained view of this data can be generated periodically by aggregation and transmitted to the monitoring server. Filtering is a different function applied to the data, where only the values of interest are selected to transmission. Other different functions can be implemented on the collected data.

\subsection{Intrusiveness}

Resource and service monitoring can be performed through different levels of intrusiveness. The monitoring solutions can be classified as active or passive, according to the interference they inject on the metrics being monitored \citep{morton2016active}, e.g., if the process of collecting metric values changes the system load, this process is called as active. Otherwise, it is a passive process.

Different metrics can have different level of intrusiveness. Running a local agent to collect CPU metric values can interfere on the value collected, since the local agent itself will use CPU cycles to perform its actions. Otherwise, StatsD \citep{noauthor_statsd_nodate}, a network protocol, permit the collection of network metrics passively without interfering in the system's network load.

\subsection{Scalability}

Scalability is a non-functional requirement \citep{taherizadeh2017auto} that guarantees a fog monitoring solution can scale to absorb an increase on monitored nodes without relevant degradation in the overall system performance. The architectural choices, such as topology, monitoring probes, communication protocols, database for local and long-term storage are relevant to determine scalability capacity of a fog monitoring solution. The proposed taxonomy defines as ``Scalable''  each monitoring solution that was evaluated for scalability and presented evidences that the proposed system scales when needed. Otherwise, the proposal was classified as ``Not Scalable''.

\subsection{Monitoring Overhead}

In order to collect instrumentation data from the monitored subjects, transmit and store them in the orchestrator, it is expected that this process consume part of environment's computing capacity and network bandwidth. Thus, the monitoring itself can be a source of resource contention, particularly in virtualized environments, where agents and  applications running in the same execution environment compete for shared resources \citep{popiolek2012monitoring}. The higher is the amount of data being collected, the higher is the overhead caused by the monitoring process \citep{popiolek2018reducing}. 
While delivering monitoring data and functionalities to the fog service orchestrator, a proper fog monitoring solution should maintain communication and processing overhead as low as possible.  The proposed taxonomy defines as ``Low Overhead'' a fog monitoring solution that was evaluated about the overhead it injects in the system and presented the results and the evaluation scenario.

\subsection{Adaptability}

Fog infrastructure is composed of heterogeneous, resource-restricted devices connected by potentially unstable communication channels. Besides, mobility is an expected characteristic of both fog nodes and end-user devices. In such a scenario, an adaptive monitoring process is of great value. According to the overall load in the system, in the monitored node or in the monitoring server, some choices can be made to diminish the impact of the monitoring process. This adaptive behavior can be applied to the frequency of instrumentation data collection, the volume and type (instrumentation domain) of data to transmit, and to the communication model. The goal is to restrict or postpone some actions while the load in the system is high, and try to resume them as soon as possible.

\subsection{Integration}
This domain verifies if the fog monitoring solution adheres to some standardized data format. Some solutions use XML or JSON files as a basic standardized data format. In recent years OpenMetrics \citep{openMetrics} was published to try to standardize the exchanging of metric data among monitoring solutions. In the same way, OpenTelemetry \citep{openTelemetry} was proposed, based on two former standards: OpenCensus \citep{openCensus} and OpenTracing \citep{openTracing}. Other standards like OpenXTrace \citep{okanovic2016towards} were proposed, although they did not get much attention from academia and industry. 
The benefit of using such standards is to ease the data exchange and integration of different monitoring components. These components could be from different vendors, and could be on different layers of Fog Architecture (Section \ref{sec:Fog}).

\subsection{Address Orchestration Needs}
Fog service orchestration needs monitoring data to implement a proper decision-making and to integrate the several functionalities that comprises it (Section \ref{Subsec:MonitOrch}).
A fog monitoring solution that is Ready to address orchestration needs is one that is:
\begin{enumerate}
    \item lightweight (e.g., implemented with a small code footprint) and multi-platform (e.g., it provides different versions of its agents to comply with resource-restriction and heterogeneity of fog nodes);
    \item Focused on monitoring - this means it does not implement other orchestrating functionalities or at least allow to turn them off, causing a minor overhead and management effort;
    \item  Adaptive - being flexible enough to change its behavior on the fly, according to configuration changes delivered by the fog service orchestrator, a coordination and management system that has a broader and comprehensive view of fog infrastructure and running services. 
\end{enumerate}

If a monitoring solution meets at least two of these requirements and can be adapted with a reasonable effort to meet the third, it is considered as being Partially Ready to address orchestration needs.  Otherwise, it is considered Not Ready.

\section{Analysis of fog monitoring tools based on proposed taxonomy} \label{Sec:Evaluation}


In this section we present a categorization of fog monitoring solutions based on the domains and categories defined in the taxonomy so researchers and developers can easily learn about the characteristics of these solutions.
Due the relevant differences among cloud and fog computing, there was no guarantee that a monitoring solution developed to the cloud would function properly in a fog computing environment \citep{gro2018comparison}. To confirm this, some recent works analyzed and tested open source and commercial cloud monitoring solutions, e.g., Nagios\citep{nagios}, Zabbix\citep{zabbix}, DARGOS\citep{povedano2013dargos}, PCMONS\citep{de2011toward} and JCatascopia\citep{trihinas2014jcatascopia}. These solutions were confronted to fog computing requirements and challenges, and the result was that none of them is suitable for fog environments \citep{abderrahim2017holistic,taherizadeh2018monitoring, abreha2021monitoring, battula2019efficient}. To overcome the monitoring challenges, some authors proposed monitoring solutions and architectures specific to fog computing environments and related paradigms. The next subsections describe each of them, approaching their characteristics according to the domains and categories that compose the Taxonomy presented in the last section.

\subsection{FMonE}

Brandón et al. \citep{brandon2018fmone} proposed FMonE as a solution that meet the fog monitoring requirements they have described in their work. FMonE is based on Marathon \citep{noauthor_marathon_nodate}, a well-known container orchestration solution, although the paper points that another container solution could be used if it meets the requirements. 

FMonE has the purposes of performance monitoring and of supporting fault tolerance. The solution uses a centralized and hierarchical topologies to collect metric values at a periodic rate using pull and push communication models. It is prepared to monitor infrastructure, platform and service layers and implements the filtering of monitoring data.

The authors used Grid5000 testbed \citep{balouek2012adding} to simulate a fog infrastructure, using 78 VMs and setting  bandwidth and latency among them. They have evaluated the service performance in operations per second in the nodes, comparing centralized versus hierarchical architectures with the use of FMonE. The results shown the solution is scalable and had a little overhead of resource consumption, running same service with and without the FMonE agent installed on the nodes.

This solution is offered as a standalone independent monitoring framework, in which the end user directly interacts and creates monitoring workflows. In a fog service orchestration scenario, the input parameters will be provided by the Service Management module (Section \ref{sec:Fog}), according to service requirements and user needs at requesting time. Also, it is not focused on monitoring, since it is responsible to detect new nodes to monitor. Nevertheless, these two issues can be adapted with reasonable effort. It is lightweight and multi-platform, but as it is not adaptive, it does  not  address  orchestration needs. FMonE is available on Github \citep{fmone_github}.

\subsection{PyMon}

The work of Großmann and Klug \citep{grossmann2017monitoring} proposes PyMon, a resource monitoring framework for ARM based single board computers (SBC), which aims to provide host and container utilization data in order to support a more efficient orchestration of containerized services.
PyMon is built as an extension of Monit \citep{noauthor_monit_nodate}, a monitoring tool capable of inspecting Docker containers. Monit is a lightweight open-source tool that is developed for monitoring Unix based systems. It is delivered by Docker images and they run on architectures supporting Docker, namely x86\_64, ARM and AARCH64.

Pymon has the purpose of performance monitoring. The solution uses a centralized topology to collect metric values from IoT devices at a periodic rate using push communication model. 
It is prepared to monitor infrastructure and platform layers and implements the aggregation of monitoring data. The received data is stored in a PostgreSQL database and can be displayed through a web interface.

Pymon is a simple, lightweight and multi-platform fog monitoring solution with low overhead of resource consumption and developed to run on SBCs. Although focused on monitoring, its feature set is not enough to meet the requirements listed on Section \ref{subsec:requirements}, supporting only Locality, as it has a local agent to collect metrics. It is not adaptive, since presents low flexibility in terms of available communication models, data transmission frequency and on-the-fly configuration changing, limiting the monitoring scenarios that are supported by it. Besides, scalability of PyMon was not evaluated. Due to these limitations, PyMon does not address orchestration needs. PyMon is available on Github \citep{pymon_github}.

\subsection{Prometheus Stack}
Prometheus \citep{prometheus} stack is a monitoring system built with the integration of Prometheus server and other complementary open-source components, like metric exporters and dashboards. For collection of monitoring data, its developers provide the tool Node Exporter, which collects metrics on Unix based systems. For container monitoring, CAdvisor \citep{noauthor_cadvisor_nodate}, developed by Google, is the chosen tool. It provides information about resource usage of host and running containers on a single machine. 

Prometheus stack has the purpose of performance monitoring. The solution uses a centralized topology to collect metric values at a periodic rate using pull communication model. 
It is prepared to monitor infrastructure and platform layers and implements the aggregation of monitoring data. 

According to Gro{\ss}mann and Klug \citep{gro2018comparison}, Prometheus stack showed a good adaptability supported by a loose coupling of its software components. Thus, it is not a complex task to modify parts of the framework to better adapt it to new scenarios. Besides, the Prometheus server has compatibility to many exporters and this can be used to collect metrics from databases, web servers and services. 

The default communication model used in Prometheus is Pull, but it also supports Push through an already implemented gateway \citep{trakadas2018scalable}. Metric exchange format defined by Prometheus was the basis to create OpenMetrics \citep{openMetrics}. So, Prometheus stack is a scalable, focused, standardized, adaptive and comprehensive monitoring solution. Although, there is no evaluation about its overall overhead on the system, its flexibility and high modularity guarantee that it is possible to configure and use it in a scenario of low overhead. So, Prometheus Stack is classified as totally addressing the orchestration needs. Prometheus Stack is available on GitHub \citep{prometheusstack_github}.

\subsection{Osmotic Monitoring}

Souza et al. \citep{souza2018osmotic} proposed a tool to monitor microservices deployed in an Osmotic computing environment, i.e., a fog-cloud environment that allows a bidirectional flow of microservices. It is an extension of CLAMBS \citep{alhamazani2015cross}, a microservice multi-cloud performance monitoring tool.

Osmotic Monitoring has the purpose of performance monitoring. The solution uses a centralized topology to collect metric values at a periodic rate using push communication model. 
It is prepared to monitor only the infrastructure layers and do not implement any data processing feature. 

The solution requires an agent on each IoT device that send data to the Manager, a component that runs in the cloud. 
The evaluations were about CPU, latency, and memory usage in six different scenarios: three on the cloud and three on the fog. Among the scenarios, variations on using only one container to hold more than one microservice and one container by microservice were compared.  Although Osmotic computing  allows a bidirectional migration of microservices between cloud and fog, the experiments were made with them standing on fixed  positions.

The work neither approach the overhead or the scalability of the proposal. It is specific to the scenario of Osmotic Computing and it is not adaptive. Its server is located on the cloud and service management and resource management are not fully separated from the monitoring functionality so it does not address the needs of a fog orchestrator.

\subsection{Support and Confidence (SCB) based Monitoring}
The work \citep{battula2019efficient} proposed a Support and Confidence (SCB) based technique, aiming to optimize the resource usage in the resource monitoring service. SCB is based on predicting confidence of each fog device, based on its historical data.
Adapting from cloud propositions, the work proposed algorithms to develop Push, Pull and Hybrid communication models, evaluated these models on a prototype build on java and compared them to the SCB based approach.

SCB has the purpose of performance monitoring. The solution uses a hierarchical topology to collect metric values at a periodic rate or triggered by events, using push, pull and hybrid communication models. 
It is prepared to monitor only the infrastructure layers and do not implement any data processing feature. 

Its performance is evaluated by analyzing a real-time traffic use case in a fog emulator and the results are compared with traditional distributed computing techniques. Results show that the proposed technique consumes fewer resources when compared to conventional resource monitoring approaches, resulting in a low system overhead and good scalability. 
It is adaptive and considers that other service management and resource management processes exist and are dependent of monitoring data and features. Thus, this proposal addresses fog orchestration needs.

\subsection{Monitoring for Fog and Mobile Cloud}
The authors of \citep{mourlin2018monitoring} proposed a MCC/Cloudlet-based architecture, composed of distributed cloudlets within multiple locations to support mobile devices using cloud services. Despite the differences between cloudlets and fog nodes, where the former has a higher computing capacity and is called cloud-in-a-box, the proposed architecture and monitoring tools chosen can be used in a fog environment with small adaptations and under certain conditions, e.g., in an IIoT use case where the fog nodes are resource richer.
The cloudlets are connected to a monitoring system and their solution is based on IEEE 1451 to communicate the sensors in a Wireless Sensor Network (WSN) and the cloudlets. It uses Virtual Device Representation (VDR), a ``digital twin'' of a device that is localized on the cloudlet. Sensu \citep{porter2016sensu}, a flexible monitoring framework, is used to implement monitoring functions and Graphite and Grafana to data storage and visualization, respectively.

The solution has the purpose of performance monitoring. It uses a centralized topology to collect metric values and logs at a periodic rate, using push communication model. 
It is prepared to monitor infrastructure, platform and service layers and implements data filtering and aggregation. 

No evaluation was made about the overhead the proposal causes on the system, but its scalability is a proven characteristic of Sensu. Sensu delivers a comprehensive framework for monitoring data processing, besides event management. Due to be standardized and adaptive, providing high customization possibility and several platform implementations, this solution addresses the fog service orchestration needs. 

\subsection{FogMon}
The works \citep{brogi2019measuring} and \citep{forti2021lightweight} proposed FogMon, a lightweight hierarchical P2P monitoring tool, based on an agent that runs on every fog node, measuring and reporting about the use of hardware resources and end-to-end network QoS between those nodes. It also detects automatically IoT devices attached to the nodes. 
FogMon adaptively and automatically modifies its P2P overlay based on current network conditions to maintain monitoring accuracy and scalability invariants. It can handle nodes that leave and join the network and relies on differential monitoring updates to reduce the overall network overhead.

FogMon has the purposes of performance monitoring and of supporting fault tolerance. The solution uses a hierarchical topology to collect metric values at a periodic rate using pull and push communication models. It is prepared to monitor only the infrastructure layer and implements the aggregation of monitoring data.

The authors have developed a prototype that was evaluated in a real testbed. The evaluation measured FogMon's footprint (usage of CPU/RAM and network by the FogMon agent) verifying it is lightweight, non-intrusive and scalable. System adaptation based on configuration changing was also confirmed. Due to not being focused only in monitoring, this solution is classified as partially addressing fog service orchestration needs. FogMon is available on Github \citep{fogmon_github}.

A most recent work, named Adaptive FogMon \citep{colombo2022towards}, added another layer of adaptivity to FogMon \citep{fogmon_github}. The authors implemented a lightweight rule-base expert system that exploits the monitoring collected data in order to adjust node's behaviour. It aims to reduce resource usage and power consumption on the node. It implemented two countermeasures that are activated based on the rule system when needed: i) Indicators Selection, which reduces the number of metrics being collected and ii) Rate, which modifies the frequency of metric delivering. When compared to FogMon, Adaptive FogMon saved energy and resources at a cost of a larger memory usage. As it is based on FogMon, which already has an adaptivity feature, only FogMon will compose Table \ref{tab:Taxonomy}. Adaptive FogMon is also available on Github \citep{adaptivefogMon_github}.

\subsection{Switch}

Taherizadeh et al. \citep{taherizadeh2018capillary} presented a capillary distributed computing architecture. It follows a reference model for autonomic services called Monitor-Analyze-Plan-Execute over a shared Knowledge  (MAPE-K) \citep{arcaini2015modeling}.
The proposed architecture includes a monitoring system, named Switch \citep{noauthor_switch_nodate},

Switch has the purpose of performance monitoring. The solution uses a centralized topology to collect metric values at a periodic rate using push communication model. It is prepared to monitor infrastructure, platform and service layers and implements event management upon monitoring data.

The collector agents are developed using the non-intrusive StatsD protocol \citep{noauthor_statsd_nodate}. The server stores received data in Cassandra \citep{noauthor_cassandra_nodate}, a free, open-source time series database (TSDB).
The monitoring system uses Docker containers, being a lightweight and multi-platform option.

Monitoring agents, server and other components like an alarm-trigger, responsible for analyzing monitoring data and creating events and notifications, were proposed in authors' previous work \citep{taherizadeh2019dynamic}.

Although it is a solution that is scalable, adaptive and with low overhead, it offers low flexibility in terms of communication models and frequency of data collection. Therefore it is considered as partially addressing fog service orchestration needs. Switch monitoring system is available on GitHub \citep{noauthor_switch_nodate}.

\subsection{TEEMon}

Trusted Execution Environment (TEE) is a promising approach to address security challenges in distributed environments, like fog computing. TEEs improves the confidentiality and integrity of application code and data even against privileged attackers with root and physical access by providing an isolated secure memory area. 

TEEMon \citep{krahn2020teemon} is a continuous performance monitoring and analysis tool for TEE-based applications.
It provides performance metrics during runtime and assists the analysis of identifying causes of performance issues. It integrates with Prometheus and Grafana, well-known monitoring  open-source tools, aiming for a holistic monitoring solution, particularly optimized for systems deployed through Docker containers or Kubernetes \cite{noauthor_kubernetes_nodate}. 
TEEMon consists of four core components:
1-Performance metrics exporters; 2-Performance metrics aggregator; 3-Performance metrics analyzer, and Performance metric visualizer. 

TEEMon has the purpose of performance monitoring. The solution uses a centralized topology to collect metric values at a periodic rate or triggered by events, using pull communication model. It is prepared to monitor only the infrastructure layer and implements filtering and aggregation of monitoring data.

It is lightweight and multi-platform, since can work with TEEs from many vendors and can monitor Docker-based applications. It is scalable, standardized, presents low overhead, and it is focused on monitoring. But as it is not adaptive, it was classified as partially addressing fog service orchestration needs. TEEMon is available on GitHub \citep{teemon_github}.

\subsection{Rule Based}

The authors of \citep{bali2019rule} proposed a monitoring system targeted to systems based on container technology. It leverages the use of rules for evaluating the importance of metrics. 
It is composed of workers and master nodes. The worker node is composed of three modules: Metrics collector, Rules updater, and Analyzer. Analyzer is responsible for processing the data collected by the Metrics Collector and evaluate them according to the current rules set. This evaluation will support the decision if that metrics set should be transmitted to the master node, for further processing and storage. Rules Updater is responsible to manage the set of rules that are updated according to the scenarios observed by the master node when analyzing the metrics sent by several workers nodes. 

Rule based monitoring has the purpose of performance monitoring. The solution uses a centralized topology to collect metric values at a periodic rate or triggered by events, using push communication model. It is prepared to monitor infrastructure and platform layers and implements filtering of monitoring data.

This proposal was neither evaluated for scalability or the overhead it injects in the system. Although very simplistic in terms of monitoring functionalities, it is lightweight, multi-platform, focused on monitoring and adaptive. Thus it addresses fog service orchestration needs.

 

\setlength\tabcolsep{2pt}

\begin{sidewaystable*}[]

\centering
\caption{Comparative analysis of fog monitoring solutions.}
\vspace{8pt}
\label{tab:Taxonomy}
\footnotesize

\begin{tabularx}{1\textheight}{lcccccccccc}
 \hline
 \textbf{Paper}               & 
\begin{tabular}[c]{@{}c@{}}\textbf{FMonE}\\
\end{tabular} & 
\begin{tabular}[c]{@{}c@{}}\textbf{PyMon}\\
\end{tabular} & 
\begin{tabular}[c]{@{}c@{}}\textbf{Prometheus}\\ 
\end{tabular} & 
\begin{tabular}[c]{@{}c@{}}\textbf{Osmotic}\\
\end{tabular} & 
\begin{tabular}[c]{@{}c@{}}\textbf{SCB}\\
\end{tabular} & 
\begin{tabular}[c]{@{}c@{}}\textbf{Mobile}\\ 
\end{tabular} & 
\begin{tabular}[c]{@{}c@{}}\textbf{FogMon}\\
\end{tabular} & 
\begin{tabular}[c]{@{}c@{}}\textbf{Switch}\\
\end{tabular} &
\begin{tabular}[c]{@{}c@{}}\textbf{TEEMon}\\
\end{tabular} &
\begin{tabular}[c]{@{}c@{}}\textbf{Rule Based}\\
\end{tabular} 

\\ \hline

\textbf{Purposes}            & 
\begin{tabular}[l]{@{}l@{}}Performance\\Fault Tolerance\end{tabular} & Performance & Performance & Performance  & 
\begin{tabular}[l]{@{}l@{}}Performance\\Resource Usage\\Fault Tolerance\end{tabular} & 
\begin{tabular}[l]{@{}l@{}}Performance\\Resource Usage\end{tabular} & 
\begin{tabular}[l]{@{}l@{}}Performance\\Fault Tolerance \end{tabular}& Performance & Performance & Performance     \\ \hline
\textbf{Topology}            & 
\begin{tabular}[l]{@{}l@{}}Centralized\\Hierarchical\end{tabular} & Centralized & Centralized & Centralized & Hierarchical & Centralized & Hierarchical & Centralized & Centralized & Centralized \\ \hline
\begin{tabular}[l]{@{}l@{}}\textbf{Communication}\\\textbf{Model}\end{tabular} & Push/Pull & Push & Pull & Push & Push/Pull/Hybrid & Push & Push/Pull & Push & Pull & Push    \\ \hline
\textbf{Frequency}           & Periodic & Periodic & Periodic & Periodic &  
\begin{tabular}[l]{@{}l@{}}Event\\Periodic\end{tabular} & Periodic & Periodic &  Periodic & 
\begin{tabular}[l]{@{}l@{}}Event\\Periodic\end{tabular} & 
\begin{tabular}[l]{@{}l@{}}Event\\Periodic\end{tabular} \\ \hline
\begin{tabular}[c]{@{}l@{}}\textbf{Monitoring}\\\textbf{Layers}\end{tabular}              & 
\begin{tabular}[l]{@{}l@{}}Infra\\Platform\\Service\end{tabular} & 
\begin{tabular}[l]{@{}l@{}}Infra\\Platform\end{tabular} & 
\begin{tabular}[l]{@{}l@{}}Infra\\Platform\end{tabular} & 
\begin{tabular}[l]{@{}l@{}}Infra\\Platform\end{tabular} & Infra & 
\begin{tabular}[l]{@{}l@{}}Infra\\Platform\\Service\end{tabular} & Infra & 
\begin{tabular}[l]{@{}l@{}}Infra\\Platform\\Service\end{tabular} & Infra & 
\begin{tabular}[l]{@{}l@{}}Infra\\Platform\end{tabular}\\ \hline
\begin{tabular}[c]{@{}l@{}}\textbf{Instrumentation}\\\textbf{Domains}\end{tabular}     & Metrics & Metrics & Metrics & Metrics & Metrics & 
\begin{tabular}[l]{@{}l@{}}Metrics\\Logs\end{tabular} & Metrics & Metrics  & Metrics & Metrics\\ \hline
\begin{tabular}[c]{@{}l@{}}\textbf{Data}\\\textbf{Processing}\end{tabular}     & Filtering & Aggregation & Aggregation & \xmark & \xmark &  \begin{tabular}[l]{@{}l@{}}Filtering\\Aggregation\end{tabular} & Aggregation & 
\begin{tabular}[l]{@{}l@{}}Events\\Management\end{tabular}  & 
\begin{tabular}[l]{@{}l@{}}Filtering\\Aggregation\end{tabular}& Filtering\\ \hline
\textbf{Intrusiveness}       & Active & Active & Active & Active & \xmark & Active & 
\begin{tabular}[l]{@{}l@{}}Active\\Passive\end{tabular} & 
\begin{tabular}[l]{@{}l@{}}Active\\Passive\end{tabular} & Active & Active  \\ \hline
\textbf{Scalability}         & \cmark & \xmark & \cmark & \xmark & \cmark & \cmark & \cmark & \xmark & \xmark & \xmark \\ \hline
\begin{tabular}[c]{@{}l@{}}\textbf{Monitoring}\\
\textbf{Overhead}\end{tabular}   & \cmark & \cmark & \xmark & \xmark & \cmark & \xmark & \cmark & \xmark & \cmark & \cmark   \\ \hline
\textbf{Adaptability}           & \xmark & \xmark & \cmark & \xmark & \cmark & \cmark & \cmark & \cmark  & \xmark & \cmark\\ \hline
\textbf{Standardized}     & \xmark & \xmark & \cmark & \xmark & \xmark & \cmark & \xmark & \xmark  & \cmark & X\\ \hline
\begin{tabular}[c]{@{}l@{}}\textbf{Orchestration}\\ \textbf{Needs}\end{tabular} & \xmark & \xmark & \cmark & \xmark & \cmark &  \cmark & $\partial$ & $\partial$ & $\partial$ &  \cmark\\ \hline     
\multicolumn{11}{l}{\begin{tabular}[c]{@{}l@{}}\textbf{\cmark} denotes that the item is implemented/addressed;\\ 
\textbf{$\partial$} denotes the item is partially addressed;\\
\textbf{\xmark} denotes that the item is not implemented/addressed;
\end{tabular}} 
\end{tabularx}

\end{sidewaystable*}

\subsection{Discussion}

To allow researchers easily find the characteristics of each analyzed fog monitoring proposal, a detailed classification based on the previously defined taxonomy is presented in Table \ref{tab:Taxonomy}.
The papers are presented in the same order they have appeared in this Section. 
By analyzing Table \ref{tab:Taxonomy} it is possible to see that the most common characteristics of available fog monitoring solutions are for purpose of performance monitoring, a centralized topology with the monitoring agent collecting infrastructure metric values, in an active manner, sending them periodically to monitoring server using push communication model and partially addressing fog service orchestration needs.

There are possible billing models such as consumption-based, where users are billed per usage, or subscription-based, where users  pay  a  fixed  monthly  rate  and  can  use  the  fog  on  a network-wide basis \citep{bittencourt2015towards}. But pricing and billing remain a challenge in terms of sustaining a commercial ecosystem of value-added services, as the business model is still not clear \citep{Yi2015}, and due to the lack of fog providers \citep{Mahmud16}. These arguments can explain why none of the works had the purpose of generating accurate bills, although monitoring plays an important role in this area.

Although centralized is the most used control topology among analyzed papers, some of them did not verify solution's scalability \citep{grossmann2017monitoring, souza2018osmotic, taherizadeh2018capillary} and this can increase the risk of failure by resource exhaustion on the monitoring server in case of huge volume of monitored nodes or in case of a bursty scenario. 
Fog computing is a distributed  paradigm. One may think that the proper control topology would also be a distributed scheme, but this assumption is not confirmed by the available fog monitoring proposals. 
According to Ward et al. \citep{ward2014observing}, distributed topologies have inherent scalability improvements over centralized ones. But it brings along a set of different challenges including system initialization (bootstrap problem), node lookup process and data replication. Monitoring solutions with distributed topologies should first overcome these challenges and risks, but they may become slower and more cumbersome than centralized solutions.

Most of the analyzed works collect monitoring data periodically, and amongst them, some proposals use only the Push method as communication model \citep{souza2018osmotic, mourlin2018monitoring, taherizadeh2018capillary, bali2019rule}. This combination may lead to scenarios where a large amount of data can be injected into the system, depending on the number of devices and services being monitored. In those scenarios, communication channels and monitoring server can become overloaded, potentially causing inefficacy, data loss or even unavailability of monitoring system \citep{krahn2020teemon}. It is fundamental to verify if those proposals are scalable and if the overhead they inject into the system is the lowest possible to permit them to cope with the aforementioned scenarios.

Observability and the management of instrumentation domains (Section \ref{Sec:Monitoring}) other than metrics are recent themes in the monitoring context. Only one out of ten analyzed proposals collect logs \citep{mourlin2018monitoring}. None of them collect traces. Logs and traces are related to white-box monitoring and can help assess the internals of the services, aiming anticipating malfunctioning, debugging of already detected problems and double-checking that everything is working properly in specific moments, e.g., after updating the service with a new version \citep{kaldor2017canopy}. Unify the data life-cycle management of instrumentation domains (metrics, logs and traces) can lower the effort of maintaining multiple data pipelines \citep{karumuri2021towards}. On the other hand, it is critical to consider the heterogeneous nature of instrumentation data in terms of frequency of generation, data volume and consumption (what determines the kind of storage and user queries), balancing the benefits of using a same data management engine with the risks of a more complex and fragile monitoring system \citep{ewaschuk2016monitoring}.

Regarding the needs of a fog service orchestrator, whose requirements were described on Section \ref{Subsec:MonitOrch}, some analyzed proposals have fully met them: \citep{prometheus, mourlin2018monitoring, battula2019efficient, bali2019rule}. This is the answer to the second research question of our review: ``RQ2 - What monitoring solutions are prepared to compose a fog service orchestrator?''. These proposals are lightweight and multi-platform, focused on monitoring and accept on-the-fly configuration changing, allowing the management from the fog service orchestrator. Nevertheless, Switch \citep{taherizadeh2019dynamic, noauthor_switch_nodate} and FogMon \citep{brogi2019measuring, forti2021lightweight, colombo2022towards}, although categorized as partially addressing orchestration needs, should be considered and be monitored since they can evolve and  easily change their categorization.

In order to help the researchers easily find the strengths and weaknesses found in each analyzed solution, we summarize this information in Table \ref{Tab:Strengths}.

\begin{table}[h] {
	\footnotesize
	\centering
	\caption{Strengths and weaknesses of analyzed fog monitoring solutions.}
	\vspace{8pt}

\resizebox{\textwidth}{!}{%
\begin{tabular}{@{}llcll@{}}
\toprule
\multicolumn{1}{l}{\textbf{Paper}} &
  \multicolumn{1}{c}{\textbf{Year}} &
  \multicolumn{1}{c}{\textbf{Open-source}} &
  \multicolumn{1}{c}{\textbf{Strengths}} &
  \multicolumn{1}{c}{\textbf{Weaknesses}} \\ \midrule
FMonE      & 2018 & Yes & Modularity, flexible architecture, real testbed               & Not adaptive, specific for Marathon                 \\ \midrule
Pymon      & 2017 & Yes & Simple, focused on monitoring, lightwheighted                 & Low feature set                                     \\ \midrule
Prometheus & 2018 & Yes & Multiple evaluations, full featured, modularity, standardized &                                                     \\ \midrule
Osmotic    & 2018 & No  &                                                               & Limited feature set,  osmotic only,   may not scale \\ \midrule
SCB        & 2019 & No  & Simple, focused on monitoring, lightwheighted                 & Infrastructure only, no data processing             \\ \midrule
Mobile     & 2018 & No  & Derived from Sensu,                                           & Require resource-rich nodes, may not scale          \\ \midrule
Fogmon     & 2022 & Yes & Adaptivity, close to be ready for orchestration               & Infrastructure only                                 \\ \midrule
Switch     & 2018 & Yes & Close to be ready for orchestration                           & Low feature set, may not scale                      \\ \midrule
TEEMon     & 2020 & Yes & Same as Prometheus                                            & May not scale                                       \\ \midrule
Rule Based & 2019 & No  & Adaptivity, flexible configuration                            & Low feature set, may not scale                      \\ \bottomrule
\end{tabular}%
}
\label{Tab:Strengths}
}
\end{table}

\section{Related Work} \label{Sec:Related}
This Section analyzes other surveys and works that proposed  monitoring taxonomies in the areas of cloud computing, fog computing and related paradigms and compares them with this work.

Some papers have presented monitoring taxonomies on cloud computing \citep{aceto2013cloud,ward2014observing,syed2017cloud, da2019survey}.  In the work of Ward and Baker \citep{ward2014observing} the authors surveyed monitoring tools and derived a taxonomy to classify them. They defined the following domains: architecture, communication mechanism, collection mechanism, environment monitored and use-case. The work also enumerate the challenges of cloud monitoring and the requirements a proper tool must have to deal with them. Finally they classified each tool by the domains and described how each one deal with the challenges presented and whether they meet the requirements.

Syed et al. \citep{syed2017cloud} surveyed the literature and created a cloud monitoring taxonomy with the following domains: monitoring purposes, communication models, overhead of monitoring system, scalability and architectural design. They included yet domains related to the business and operational models, e.g., monitoring perspective (user x provider), type of cloud (public x private) and license (open-source x commercial).

In \citep{da2019survey}, the authors combined system monitoring, resource management and load prediction, which they named as global management view. They argued that the three areas have a high correlation regarding the performance of enterprise systems. They proposed a taxonomy for each area independently, but discussed about their interconnections. They categorized only cloud and on-premises proposals with the defined taxonomy.

Despite the relevant differences between cloud computing and fog computing, some domains remain valid and are referenced on fog computing proposals. In the paper \citep{abderrahim2017holistic}, the authors described what key properties a fog monitoring system must have to better deal with fog characteristics. Besides, it categorized monitoring architectures according to their topology -- centralized, hierarchical and P2P -- and to their functional decomposition granularity (no decomposition, basic decomposition and fine-grained decomposition). The authors analyzed several cloud and on-premises monitoring solutions from the literature and categorized them  according to the properties and architectures described in the work.

In the work of Taherizadeh et al. \citep{taherizadeh2018monitoring}, the authors depicted fog monitoring process as being a solution that has to deal with four levels:  VMs, containers, network connections and applications. The paper reviewed the literature, described and compared proposals found that have addressed each monitoring level. The authors proposed a taxonomy of functional and non-functional requirements of fog  monitoring solution and compared cloud monitoring tools to the requirements. They concluded that none of the existing solutions fully attend the fog monitoring requirements, so it is necessary to adapt the existing solutions or develop a new one. No implementation nor evaluation is provided.

Abreha et al. \citep{abreha2021monitoring} presented a taxonomy of fog monitoring solutions with three domains:  Architecture,  Requirements and  Design Parameters. Architecture is divided into the same categories defined by Abderrahim et al. \citep{abderrahim2017holistic}: Architectural Models and Functional Decomposition. Requirements is divided similarly to the work of Taherizadeh et al. \citep{taherizadeh2018monitoring}: Functional and Non-functional requirements. Lastly, Design Parameters is divided into  Topology, Data, Frequency Sampling and Network Bandwidth. The authors analyzed mainly cloud monitoring solutions, concluding they are not suitable for fog environments. They analyzed only two specific fog monitoring solutions.

Unlike the previously described works, this paper created a taxonomy of fog monitoring solutions covering more domains and categories, giving a more detailed view of their properties and behavior. Recent subjects were brought to the light, like the Observability and Instrumentation domains, updating the discussion in the field.
In addition, this work also addressed the orchestration needs that a fog monitoring solution must provide to properly compose a fog management system. Also, due to the relevant differences between cloud and fog paradigms, only fog proposals were considered and this work analyzed ten of them, giving the researchers access to a more updated and comprehensive knowledge base. Table \ref{Tab:RelatedWork} compares the related work presented in this section with the goals of this paper.

\setlength\tabcolsep{4pt}
\begin{table*}[!ht]
	\footnotesize
	\centering
	\caption{Related work comparison.}
	\vspace{8pt}
\begin{tabular}{ccccccc}
\hline
\begin{sideways}\textbf{Paper}\end{sideways}  & \begin{sideways}\textbf{Year}\end{sideways} & \begin{sideways}\textbf{\begin{tabular}[c]{@{}l@{}}Focus on Fog\\ Computing\end{tabular}}\end{sideways} & \begin{sideways}\textbf{\begin{tabular}[c]{@{}c@{}}Number of Fog\\Solutions Analyzed\end{tabular}}\end{sideways} & \begin{sideways}\textbf{\begin{tabular}[c]{@{}c@{}}Proposed \\Taxonomy\end{tabular}}\end{sideways} & \begin{sideways}\textbf{\begin{tabular}[c]{@{}c@{}}Addressed\\ Orchestration\\Challenges\end{tabular}}
\end{sideways} & \begin{sideways} \textbf{\begin{tabular}[c]{@{}c@{}}Addressed\\ Observability and\\ Instrumentation\\ Domains\end{tabular}}\end{sideways} \\ \hline
\citep{aceto2013cloud} & 2013 & \xmark & 0 & \cmark & \xmark & \xmark \\ \hline
\citep{ward2014observing} & 2014  &  \xmark & 0  & \cmark & \xmark & \xmark \\ \hline
\citep{syed2017cloud} & 2017  & \xmark  & 0  & \cmark & \xmark & \xmark \\ \hline
\citep{abderrahim2017holistic} & 2017  & \cmark & 0  & \xmark & \xmark & \xmark\\ \hline
\citep{taherizadeh2018monitoring} & 2018  & \cmark & 0 & \cmark & $\partial$ & \xmark \\ \hline
\citep{da2019survey} & 2019  &  \xmark & 0   & \cmark & \xmark & \xmark \\ \hline
\citep{abreha2021monitoring} & 2021  & \cmark & 2  & \cmark & $\partial$ & \xmark \\ \hline
\textbf{This Work} & 2022  & \cmark & 10  & \cmark & \cmark & \cmark   \\ \hline
\multicolumn{7}{l}{\begin{tabular}[c]{@{}l@{}}\cmark denotes comprehensive discussion about the item\\
$\partial$ denotes partial or superficial discussion about the item\\
\textbf{\xmark} denotes that the item is not implemented/addressed\end{tabular}} 
\end{tabular}
\label{Tab:RelatedWork}
\end{table*}

\section{Open Challenges} 
\label{Sec:Desafios}

This Section answer the third research question of this review: ``RQ3 - What are the challenges that still need attention from the academia?''.  In the next paragraphs we discuss about some of these challenges, relate them to the categories in the taxonomy and present some possible directions to overcome them. 

This work contextualized the high importance of monitoring in an fog service orchestration scenario. Although there are in the literature dozens of fog orchestration's proposals \citep{costa2021orchestration}, most of them have assumed that a monitoring process was already available and did not detail its requirements, architecture, tools and technical properties \citep{battula2019efficient}. To fill this gap, this work analyzed fog monitoring solutions to identify their main characteristics and to verify whether they address the orchestration's needs. As a last result from this research, we identified some challenges in this scenario and present and discuss them in this section, as follow:

\begin{itemize}
    \item Security and Privacy - Although fog service orchestrator should deliver a Security functionality, as seen on Figure \ref{fig:architecture}, it will take care of architectural security, e.g., available security standards concerning cryptography and communication channels. But as fog nodes are potentially resource-restricted, this characteristic may limit security tools and techniques that could be used, and this can make easier to an attacker to hack into the client software. So, access control, data encryption, contextual integrity and isolation mechanisms over sensitive data should be analyzed to prevent security breaches \citep{petrakis2018internet}. Viejo et al. \citep{viejo2020secure} proposed a two-protocol process of securing transmission of monitored data in a centralized topology with push communication model. First protocol identifies the node hierarchy from the client to the server and the second secure the data stream with a lightweight symmetric cryptographic protocol. 

\item Management of Stored Monitoring Data \citep{syed2017cloud} -  Deliver all monitoring collected data may congest the network and overburden the server. Store all data locally is unfeasible due to storage restrictions and the orchestrator's necessity for fast access to these data for decision-making \citep{gro2018comparison}. So different strategies should be used to balance risks and benefits. Internet of things as a service (iTaaS) \citep{petrakis2018internet} is a framework that supports data handling with low bandwidth usage. Data filtering, local data caching, postponed data uploading and data synchronization with the server are available strategies. Although developed to cloud scenario, the techniques used can be applied to fog monitoring solutions. Another strategy is the reduction in amount of metrics being monitored without loss of accuracy related to performance monitoring of resources and services. With the use of linear correlation and hierarchical clustering analysis, the authors of \citep{popiolek2021low} proposed an approach that automatically calculates correlations between available metrics, allowing the reduction of monitoring data dimensionality. Besides reducing data management effort, a local mechanism to decide when it is the appropriate time to mitigate risks is of great value. Anagnostopoulos and Kolomvatsos \citep{anagnostopoulos2019intelligent} proposed such a mechanism, with the use of Optimal Stopping Theory \citep{peskir2006optimal}, to permit that the monitoring tool on the fog node track QoS measurements and signalize orchestrator that an action is needed.

\item Heterogeneity and High Distribution -  SBCs (e.g., Raspberry Pi, Beagle board, etc.) are cheap, flexible and easy to use and integrate to a fog environment \citep{grossmann2017monitoring}, but as there is no standardized hardware to fog computing, software stacks may not be supported in all devices \citep{babu2021fog}. If a monitoring agent (or any service managed by the orchestrator) is available as Docker images, this demands that fog nodes have Docker runtime installed previously. A possible approach is the use of over-the-air (OTA) \citep{al2021faster} technique to make the bootstrap process, and to install execution environment and required software modules on the device on-the-fly, as needed. In a high distribution of fog nodes, the devices may span through different domains, and challenges like devices' clock synchronization \citep{mourlin2018monitoring, mansouri1992monitoring} must be addressed to prevent wrong results on monitoring data aggregation. Yet, monitoring federated domains can be challenging due to security, privacy and legal issues \citep{babu2021fog,abreha2021monitoring}.

\item Integration of Fog monitoring with Cloud Monitoring -  This work focused on creating a comprehensive taxonomy to evaluate and categorize fog monitoring solutions, since there are relevant differences between fog and cloud that prevent cloud proposals from being used properly in fog \citep{abderrahim2017holistic,taherizadeh2018monitoring, abreha2021monitoring, battula2019efficient}. With the use of standardized data formats, such as those defined by OpenTelemetry \citep{openTelemetry} and OpenMetrics \citep{openMetrics}, there is a possible path for the integration between the systems used in fog and in the cloud. The authors of \citep{karumuri2021towards} proposed that different components of the monitoring system be used in a distributed way among the layers of the environment. This structure could accommodate the specifics of collection and processing in the fog, with data generation on resource-limited devices, rapid transmission of alerts, and the minimum set of monitoring data that allows rapid decision-making at the edge of the network. Other processes, such as optimization (Section \ref{sec:Fog}), will request a greater volume of information (logs, traces), when possible, respecting the limits established by the SM so that SLAs are not compromised. From the storage of this data in fog, it is possible to share it with the cloud, thus integrating the systems of both layers. This approach is consistent with the recommendation made by \citep{ewaschuk2016monitoring}: ``maintaining distinct systems with clear, simple, loosely coupled points of integration is a better strategy''.

\item Lack of Comprehensive Simulation Tools to Support the Development of a State-of-the-art Fog Monitoring Solution - Fog computing simulators are systems that try to imitate the functioning of a fog environment, providing component and behavior modeling. There are dozens of fog simulators and iFogSim \citep{gupta2017ifogsim} is the most referenced one. Most of these simulators support only some basic categories of the monitoring taxonomy defined in this paper, e.g., topology, communication model, frequency, and scalability \ref{Sec:Taxonomy}, according to a survey put forward by Markus and Kertesz \citep{markus2020survey}. The authors did a detailed introduction and analysis of cloud, IoT and fog simulators and provided a comprehensive comparison of capacities and models, highlighting the different available versions and the links to the open-source repositories. To validate the features that are not  supported by a single simulator, the developers will need to use more than one simulator, which is time and effort-consuming, or develop their own extensions to better validate their use cases. Recently, Alwasel et. al has proposed a fog simulator that models Osmotic Computing and Mahmud et al. \citep{mahmud2022ifogsim2} proposed iFogSim2, a modular simulator that models service migration, dynamic distributed cluster formation, and microservice orchestration, based on real datasets. 

\end{itemize}

\section{Conclusions} \label{sec:Conclusion}

Fog computing extends cloud computing to the edge of the network, properly dealing with low-latency and real-time use cases. To provide services to the end-users and guarantee that SLA and QoE are respected, fog service orchestration coordinates the environment. Orchestration is a composition of several complementary functionalities, including Monitoring, that is specifically accountable for collecting updated status about resources and services, and for delivering them timely to support decision-making.

This paper digs deeply into the role of fog monitoring in the orchestration of a fog environment. Fog monitoring characteristics, components and requirements were presented and analyzed, and a discussion about its integration with other orchestration functionalities was presented, aiming to increase the knowledge base about this novel field of monitoring.

A taxonomy of fog monitoring solutions was created from  the most relevant domains and categories in the area. To validate the taxonomy and to offer researchers a comprehensive analysis of available fog monitoring proposals, they were analyzed and categorized by the taxonomy, showing its usefulness. Due to the relevant differences between cloud and fog computing, and based on the evidences that cloud monitoring proposals evaluated in the literature are not proper for use in the fog environment, this paper only analyzed fog monitoring proposals. Lastly, the challenges of monitoring fog infrastructures were presented, bringing some future directions of the research in this area.

\section{Acknowledgements}
This study was financed in part by the Coordenação de Aperfeiçoamento de
Pessoal de Nível Superior - Brasil (CAPES) - Finance Code 001.


\bibliography{Conference}

\end{document}